\documentclass[fleqn,usenatbib]{mnras}

\usepackage{newtxtext,newtxmath}

\usepackage[T1]{fontenc}
\usepackage[flushleft]{threeparttable}

\newcommand\chandra{{\it Chandra}}

\newcommand\kms{\ifmmode {\rm~km\ s}^{-1} \else ~km s$^{-1}$\fi}
\newcommand\Hunit{\ifmmode {\rm~km\ s}^{-1}\ {\rm Mpc}^{-1}
        \else ~km s$^{-1}$ Mpc$^{-1}$\fi}
\newcommand\ctssec{\ifmmode {\rm~count\ s}^{-1} \else ~count s$^{-1}$\fi}
\newcommand\ergsec{\ifmmode {\rm~erg\ s}^{-1} \else
        ~erg s$^{-1}$\fi}
\newcommand\funit{\ifmmode {\rm~erg\ s}^{-1}\;{\rm cm}^{-2} \else
        ~ergs s$^{-1}$ cm$^{-2}$\fi}
\newcommand\phflux{\ifmmode {\rm~photon\ s}^{-1}\;{\rm cm}^{-2}
        \else   ~photon s$^{-1}$ cm$^{-2}$\fi}
\newcommand\efluxA{\ifmmode {\rm~erg\ s}^{-1}\;{\rm cm}^{-2}\;{\rm
        \AA}^{-1} \else ~erg s$^{-1}$ cm$^{-2}$ \AA$^{-1}$\fi}
\newcommand\efluxHz{\ifmmode {\rm~erg\ s}^{-1}\;{\rm cm}^{-2}\;{\rm
        Hz}^{-1} \else ~erg s$^{-1}$ cm$^{-2}$ Hz$^{-1}$\fi}
\newcommand\cc{\ifmmode {\rm~cm}^{-3} \else cm$^{-3}$\fi}
\newcommand\FWHM{\ifmmode {\rm~FWHM} \else ${\rm~FWHM}$\fi}
\newcommand\Lsun{\ifmmode L_{\odot} \else $L_{\odot}$\fi}

\newcommand\hbeta{\ifmmode {\rm H}\beta \else H$\beta$\fi}
\newcommand\Kalpha{\ifmmode {\rm K}\alpha \else K$\alpha$\fi}
\newcommand\nh{\ifmmode N_{\rm H} \else N$_{\rm H}$\fi}

\newcommand{\Msun}{\ensuremath{\rm M_{\odot}}}
\newcommand\Zsun{\ifmmode Z_{\odot} \else $Z_{\odot}$\fi}

\usepackage{tablefootnote}

\usepackage{xcolor}

\DeclareUnicodeCharacter{2212}{\textendash}
\DeclareRobustCommand{\VAN}[3]{#2}
\let\VANthebibliography\thebibliography
\def\thebibliography{\DeclareRobustCommand{\VAN}[3]{##3}\VANthebibliography}

\usepackage{graphicx}	
\usepackage{amsmath}	
\usepackage{gensymb}
\usepackage{soul}

\title[X-ray cavity and radio mini-halo]{Evidence of sloshing-driven mini-halo formation in the cool-core cluster RXCJ1558.3-1410}

\author[Kale et al.]{Vishal S. Kale$^{1}$, S. K. Kadam$^{1}$, Sameer Salunkhe$^{2}$, S. S. Sonkamble$^{3 \textcolor{blue}{\textbf{*}}}$, N. D. Vagshette$^{4}$, \newauthor Surajit Paul$^{5}$, Ruta Kale$^{2}$, S. Ilani Loubser$^{3,6}$ and M. K. Patil$^{1}$\thanks{\color{blue}{satish04apr@gmail.com, patil@associates.iucaa.in}} \\ 
\\
$^{1}$School of Physical Sciences, Swami Ramanand Teerth Marathwada University, Nanded 431606, India\\
$^{2}$National Centre for Radio Astrophysics (NCRA), Tata Institute of Fundamental Research (TIFR), Pune 411007, India\\
$^{3}$Centre for Space Research, North-West University, Potchefstroom 2520, South Africa\\
$^{4}$Department of Physics and Electronics, Maharashtra Udayagiri Mahavidyalaya, Udgir, Dist. Latur 413517, India\\
$^{5}$Manipal Centre for Natural Sciences, Manipal Academy of Higher Education, Karnataka, Manipal 576104, India \\
$^{6}$ National Institute for Theoretical and Computational Sciences (NITheCS), Potchefstroom 2520, South Africa}

\begin{document}
\pagerange{\pageref{firstpage}--\pageref{lastpage}} \pubyear{2025}
\maketitle
\label{firstpage}

\begin{abstract}
Radio mini-halos are perplexing features, typically hosted by X-ray cool-core galaxy clusters. Understanding the connection between thermal X-ray and non-thermal radio emission is key to uncovering their origin. Here, we present a multiwavelength study of the cool-core cluster RXCJ1558.3−1410 using archival \chandra\, X-ray and wideband uGMRT radio data (Bands 3, 4 and 5). Our improved analysis confirms a previously known X-ray cavity at $\sim$36~kpc south-east of the cluster centre and we report a new cavity at $\sim$42~kpc to the north-west. These cavities suggest that the AGN provides mechanical power of $\sim$$6.0 \times 10^{44}$ erg s$^{-1}$, sufficient to offset radiative cooling in the ICM. We also detect a sharp surface brightness edge at $\sim$72~kpc south-east of the centre, characterised by a temperature jump and pressure continuity, consistent with a cold front, likely caused by gas sloshing from a minor merger. Our uGMRT images reveals an interesting diffuse emission surrounding the brightest cluster galaxy (BCG), with its edge spatially coinciding with the sloshing cold front and roughly with the cooling radius. Furthermore, a low star formation rate and uniform metal abundance up to the sloshing edge are consistent with the earlier findings of suppression of star formation and metallicity homogenisation by mixing core gas through sloshing. Finally, the spatial correlation between the mini-halo and the observed X-ray features indicates that ICM sloshing, rather than AGN feedback, plays a dominant role in powering the proposed radio mini-halo emission.

\end{abstract}



\section[1]{Introduction}

Galaxy clusters emit intense X-ray radiation, primarily due to bremsstrahlung emission from the hot intracluster medium (ICM). In the absence of a compensating heat source, this would lead to rapid cooling and inflow of gas towards the core, called cooling flow \citep{1994ARA&A..32..277F}. Past X-ray observational studies of cooling flow clusters predicted high cooling rates; however, studies revealed a significant shortfall in the amount of cool gas and much lower star formation rates than expected, pointing to a disparity now known as the ``cooling flow problem'' \citep{2006PhR...427....1P, 2007ARA&A..45..117M,McNamara_2012}.

High-resolution X-ray and spectroscopic data further confirmed the absence of gas below 1-2 keV, supporting the need for some heating process to balance cooling in cluster cores \citep{2003ApJ...590..207P}. Among various possibilities, mechanical heating from active galactic nuclei (AGN) has proven to be highly feasible. In this framework, which is known as the ``radio mode'' of AGN feedback, gas that cools near the cluster centre is drawn towards the supermassive black hole (SMBH), activating the AGN. This generates powerful outbursts in the form of relativistic jets, observed as synchrotron radio emission, that release enough energy into the ICM to offset the cooling and maintain thermal stability in the core. Observational evidence over the last two decades using high resolution X-ray data strongly supports such a feedback loop as a crucial process regulating the thermal state of cluster cores \citep{2012ARA&A..50..455F, McNamara_2012, Hlavacek_Larrondo_2022}

Among all the observed clusters, cool-core relaxed clusters are particularly noteworthy, as they often host a central AGN whose outbursts interact with the surrounding ICM \citep{2004ApJ...607..800B,2007ARA&A..45..117M, 2012ARA&A..50..455F}. This interaction of AGN and ICM leads to the formation of X-ray cavities created by highly energetic radio jets \citep{2004ApJ...607..800B,2006ApJ...652..216R,2007ARA&A..45..117M, 2007ApJ...659.1153W,Hlavacek_Larrondo_2022}. Many galaxy clusters have been thoroughly examined for X-ray cavities. Some of the well known systems that have been studied are Perseus \citep{10.1046/j.1365-8711.2003.06902.x, 2006MNRAS.373L..16F}, Hydra A \citep{2005ApJ...628..629N}, Abell 2052 \citep{2001AAS...19916110B,2009ApJ...697L..95B,2011scgg.conf....1B}, Abell 2597 \citep{2012MNRAS.424.1026T}, ZwCl 2701 \citep{2016MNRAS.461.1885V} Abell 1991 \citep{2013Ap&SS.345..183P} , Abell 3847 \citep{2017MNRAS.466.2054V}, Abell 2390,\citep{2015Ap&SS.359...61S},  3C 320 \citep{2019MNRAS.485.1981V}, Abell 2626 \citep{2019MNRAS.484.4113K}, Abell 795 \citep{Ubertosi_2021} and RXCJ0352.9+1941 \citep{2024JApA...45...23S}.

Some of these cool-core relaxed clusters display a spiral-like structure, typically linked to sloshing motion and the presence of cold fronts in the ICM \citep{2006ApJ...650..102A,2007PhR...443....1M}. These sloshing structures associated with the cold front were first detected in the cluster Abell 2142 \citep{2000ApJ...541..542M}. The cold fronts are characterised by a temperature jump with continuity in the pressure, where the brighter and denser side appears cooler, with the fainter and less dense regions as hotter \citep{2007PhR...443....1M}. These cold fronts have been studied in numerous relaxed cool-core clusters, e.g. Abell 1795 \citep{2001ApJ...562L.153M}, 3C 320 \citep{2019MNRAS.485.1981V}, Abell 3444 \citep{2024ApJ...961..133G}, Abell 2566 \citep{2024NewA..11102253K}, Abell 795 \citep{2024MNRAS.531.4060K}.

It is noteworthy that radio mini-halos, non-thermal diffuse radio structures are observed to envelop the central Brightest Cluster Galaxy (BCG) in some relaxed cool-core clusters \citep{FERETTI20041137, Feretti_2012, van_Weeren_2019}. These structures usually extend over 100–500 kpc and are typically steep spectrum in nature ($\alpha \lesssim -1$\footnote{where $\alpha$ is the spectral index defined by the relation between flux density $S$ and observing frequency $\nu$ as $S \propto \nu^{\alpha}$}). Mini-halo emission is often bounded by X-ray cold fronts associated with sloshing motions in the cluster core \citep{2008ApJ...675L...9M, 2013ApJ...762...78Z, 2024ApJ...961..133G}. However, recent sensitive observations with new generation radio telescopes have revealed emission extending beyond these cold fronts in some clusters \citep{biava_2021MNRAS.508.3995B, 2022MNRAS.512.4210R}. Two principal models explain their origin. The first is the turbulent re-acceleration scenario, wherein pre-existing mildly relativistic electrons originating from star formation or past AGN activity are re-energised by ICM turbulence triggered by gas sloshing or minor mergers that preserve the cool-core \citep{2002A&A...386..456G, 2013ApJ...762...78Z}. The second is the hadronic model, in which cosmic ray protons collide inelastically with thermal protons in the ICM, producing secondary electrons that emit synchrotron radiation \citep{pformmer_2004A&A...413...17P}. A hybrid scenario has also been proposed, combining re-acceleration of both primary and hadronically produced secondary electrons by turbulence \citep{2022MNRAS.512.4210R, 2023MNRAS.524.6052R}.

\begin{figure*}
\centering
\includegraphics[width=\textwidth, trim=0cm 0.5cm 0cm 0cm, clip]{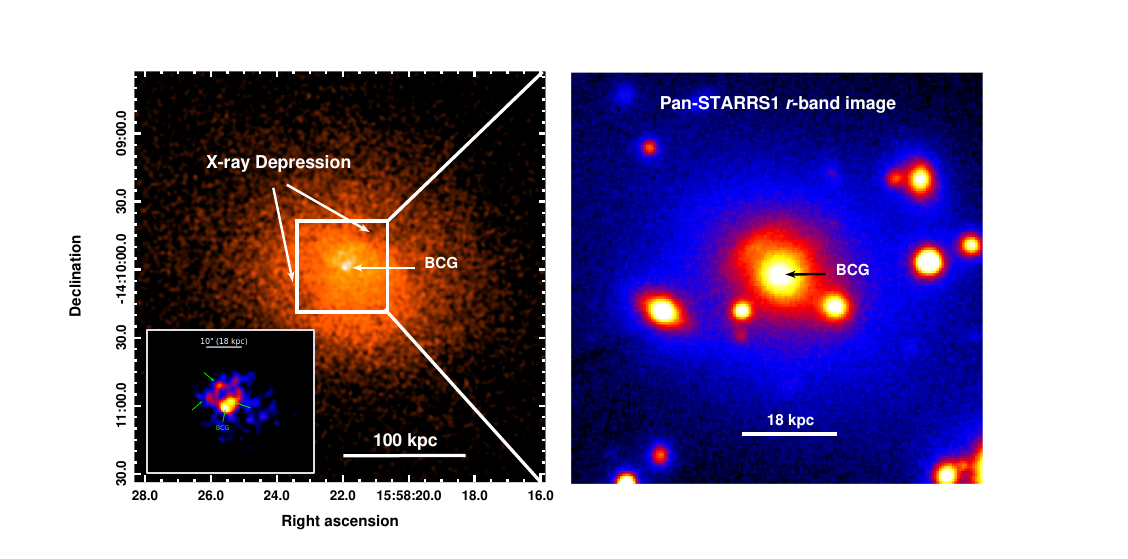}
\caption{\textit{Left panel:} 3\arcmin $\times$ 3\arcmin exposure corrected, point sources removed, background subtracted 0.5 - 3.0 keV \textit{Chandra} X-ray image of RXCJ1558 smoothed with a 2$\sigma$ Gaussian kernel. Inset close-up view of the cluster core, with X-ray substructures indicated by arrows and the BCG clearly marked. \textit{Right panel:} 40\arcsec $\times$ 40\arcsec\, Pan-STARRS1 \textit{r}-band optical image of RXCJ1558 with BCG marked by a black arrow.}
\label{raw}
\end{figure*}

RXCJ1558.3-1410 (hereafter RXCJ1558) is a known relaxed (quiescent) galaxy cluster \citep{2016MNRAS.460.1758H} with $M_{500}$ = 3.87 $\times$ 10$^{14}$ M$_{\odot}$ \citep{2011A&A...534A.109P} positioned at RA= 15h58m23.20s, DEC= -14d10m04s and located at a redshift of 0.097. The central radio galaxy PKS 1555-140 coincides with the BCG of RXCJ1558 \citep{2015MNRAS.453.1201H}. The radio study of RXCJ1558 shows the variability of the central radio-loud galaxy about 3-4 yrs at the 15 GHz radio frequency \citep{2022MNRAS.509.2869R}. The star formation rate and the ICM cooling rate in the core are 3.80$\pm$1.14 M$_{\odot}$ yr$^{-1}$ \citep{2022ApJ...940..140C} and 125$\pm$1.0 M$_{\odot}$ yr$^{-1}$  \citep{2018ApJ...858...45M} respectively. Furthermore, RXCJ1558 exhibits a steep spectrum power-law tail at low radio frequencies and hosts an unresolved central core and a double-lobed radio structure at 2 GHz and 8 GHz \citep{2015MNRAS.453.1201H}.

\citet{2013MNRAS.431.1638H} and \citet{2013MNRAS.432..530R} have reported the presence of a pair of cavities in this cluster; however, detailed analysis of the cavity energetics was not available. \cite{sonkamble2017xray} detected two cavities along the north-west to south-east direction and studied them using \textit{Chandra} X-ray data. Recently, VLBA and \chandra\, data studies on a sample of 16 galaxy clusters reported the detection of one cavity in the south-east region and evidence of a sloshing structure in this cluster \citep{2024ApJ...961..134U}.  Furthermore, two jet lobes are detected at higher radio frequencies (5 and 8.7 GHz) along a direction offset from the cavity axis. The measured jet-axis reorientation is 61.7\degr, as reported by \cite{2024ApJ...961..134U}, indicating a substantial cavity–jet misalignment likely driven by changes in AGN activity over time. However, a detailed correlation between the central AGN and the X-ray cavities has not yet been done. This paper, employing \chandra\, X-ray and radio observations using uGMRT (across Bands 3, 4, and 5) presents an in-depth analysis of the X-ray cavities, gas sloshing, and its association with the radio mini-halo.
As a result, this cluster becomes particularly interesting for further study.

The paper is organised as follows: Section~\ref{red} provides details on the \textit{Chandra} and uGMRT observations and data reduction strategies. Section~\ref{results} describes the results obtained from X-ray imaging, X-ray surface brightness, spectroscopy, and radio imaging. This section also provides a discussion on the detection of the X-ray cavities and the cold front. Section~\ref{disussions} provides a concise discussion on cavity energetics, heating and cooling mechanism of the ICM, the sloshing cold front, and its association with the radio mini-halo. Finally, Section~\ref{con} presents conclusions drawn from the present study.

Throughout the paper, we adopt the cosmological parameters $H_{0}$ = 70 km s$^{-1}$ Mpc$^{-1}$, $\Omega_{m}$ = 0.27 and $\Omega_{\Lambda}$ = 0.73; translating to a scale of 1.8 kpc per arcsec at the redshift $z$ = 0.097. All errors quoted for spectral analysis have a confidence level of 68\%.

\section[2]{Observations and the data reduction}
\label{red}
\subsection{X-ray data}
\begin{table*}
\centering
\caption{\textit{Chandra} observation details of RXCJ1558.}
\label{tab:Table_xray}   
\begin{tabular}{c c c c c c c}
\hline
ObsID & Instrument & Date Obs & Data Mode & PI & Exposure (ks) & Cleaned Exposure (ks) \\
\hline
9402 & ACIS-S & April 09, 2008 & VFAINT & JOHNSTONE & 40.06 & 39.70 \\
\hline
\end{tabular}
\end{table*}
RXCJ1558 was observed with the \textit{Chandra} X-ray telescope on April 9, 2008 (ObsID 9402) for a total exposure time of 40.06 ks (see Table~\ref{tab:Table_xray}). Level 1 event files of this cluster were reprocessed using Chandra Interactive Analysis of Observations \citep[CIAO v4.15.2,][]{2006SPIE.6270E..1VF} and CALDB\footnote{\url{https://heasarc.gsfc.nasa.gov/docs/heasarc/caldb/caldb_intro.html/}} v4.10.7 which is provided by Chandra X-ray Center (CXC) and following standard routines as discussed in \cite{2015Ap&SS.359...61S}. The reprocessing was performed using the \texttt{chandra$\_$repro} script, which applies the latest corrections for gain calibration, charge transfer inefficiency, grade filtering, and good time interval screening to generate the Level 2 event files. To eliminate intervals affected by the background flaring, light curves were extracted from the reprocessed event file and cleaned using the \texttt{lc$\_$sigma$\_$clip}\footnote{\url{https://cxc.cfa.harvard.edu/ciao/ahelp/lc_sigma_clip.html/}} tool. This procedure resulted in a final exposure time of 39.70 ks. The resultant cleaned event file contains several point sources, which were identified using the \texttt{wavdetect} algorithm with wavelet scales ranging from 1 to 16 pixels, and then visually verified. The detected point sources were removed from the event file, and the resulting holes were filled using the \texttt{dmfilth}\footnote{\url{https://cxc.cfa.harvard.edu/ciao/ahelp/dmfilth.html/}} tool by interpolating counts from the surrounding regions. This procedure preserves the underlying diffuse emission and this file was used for imaging analysis. The diffuse X-ray emission from the cluster extends across the entire CCD, no source-free region was available for local background estimation. Therefore, blank-sky background files provided by the CXC were used. The \texttt{blanksky} and \texttt{blanksky$\_$image} tools were used to identify a suitable blank-sky background dataset corresponding to the events and were normalised to match the 9 - 12 keV count rate in the event file.

\subsection{Radio data}

Archival data sets from the upgraded Giant Metrewave Radio Telescope (uGMRT; Proposal ID: 44\_039; PI: Arvind Balasubramanian) were utilised to investigate the radio properties of the galaxy cluster RXCJ1558. These observations were originally conducted for an unrelated target rather (FRB 20190520B) than RXCJ1558 serving as a secondary calibrator. The cluster was serendipitously observed in three uGMRT frequency bands: Bands 3, 4, and 5. The details of the observations are summarised in Table~\ref{tab:Table_radio}.

\begin{table*}
\centering
\caption{Radio observation details.}
\label{tab:Table_radio}   
\begin{tabular}{c c c c c}
\hline
uGMRT Band & Observing date & Bandwidth (MHz) & Number of channels & On source time (min) \\
\hline
Band 3 & June 18-19, 2023 & 300-500 & 2048 & 82 \\
Band 4 & June 16, 2023 & 550-950 & 2048 & 21 \\
Band 5 & June 20, 2023 & 1060-1460 & 2048 & 20 \\
\hline
\end{tabular}
\end{table*}

Analysing data at lower frequencies presents significant challenges, primarily due to ionospheric distortions that constrain the achievable dynamic range of the radio images. To mitigate these effects, we utilise the \texttt{Source Peeling and Atmospheric Modelling} (\texttt{SPAM}) pipeline \citep{Intema_2009A&A...501.1185I, Intema_2017A&A...598A..78I} for the uGMRT data analysis. This robust and comprehensive pipeline integrates standard data reduction procedures with direction-dependent calibration techniques, essential for correcting visibility amplitude and phase fluctuations induced by antenna primary beam variations and ionospheric irregularities. The direction-dependent calibration is facilitated by leveraging bright sources located within the primary beam (see \citealt{Intema_2009A&A...501.1185I, Intema_2017A&A...598A..78I} for further details).

\begin{figure*}
\includegraphics[width=1.0\textwidth]{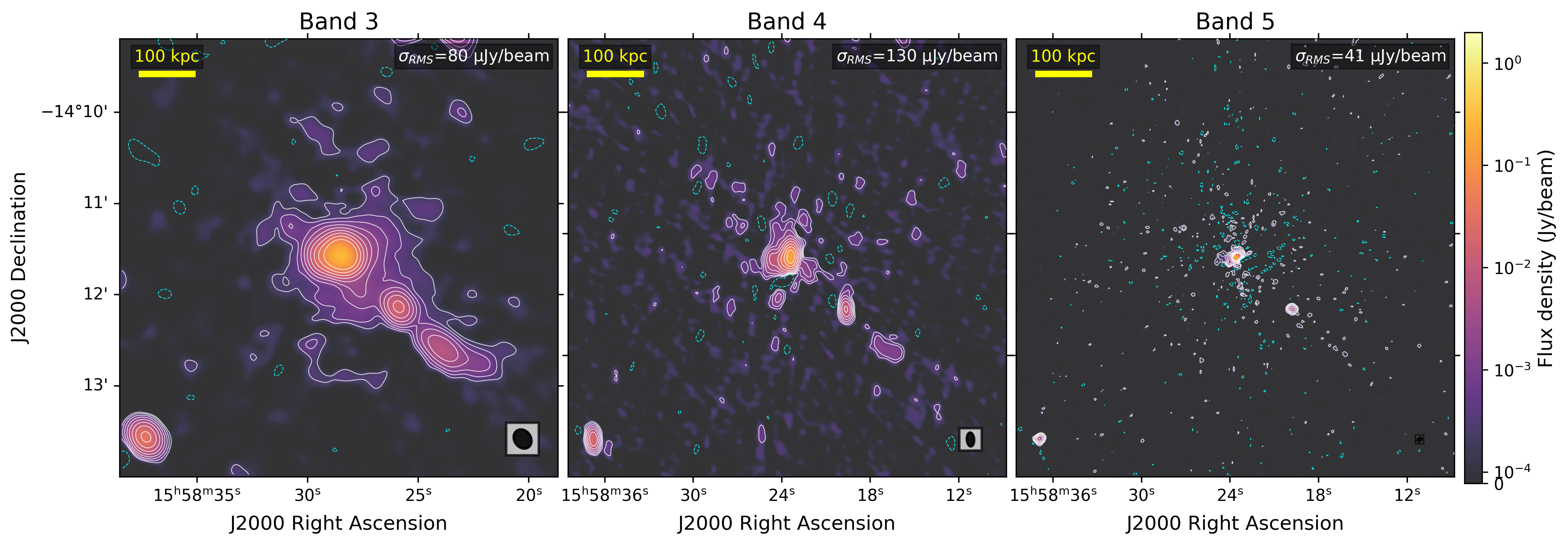}
   \caption[All band radio image] {Three-panel radio continuum observations of RXCJ1558 at different frequencies. Left to right: uGMRT Band 3 ($\sigma_{\rm rms}$ = 80 $\mu$Jy\,beam$^{-1}$), Band 4 ($\sigma_{\rm rms}$ = 130 $\mu$Jy\,beam$^{-1}$), and Band 5 ($\sigma_{\rm rms}$ = 41 $\mu$Jy\,beam$^{-1}$). The white contour levels are plotted at [3, 6, 12, 24, ...] $\times \sigma_{\rm rms}$ and a cyan dashed contour at -3$\sigma_{\rm rms}$. The synthesized beam size for each band is shown as a white box with black ellipse in the lower-right corner.}
    \label{all_band_radio}
\end{figure*}

The data in each uGMRT band were divided into six sub-bands with bandwidths of 33, 50, and 66~MHz for Bands~3, 4, and 5, respectively. Each sub-band was independently calibrated for flux density and bandpass shape using the \texttt{SPAM} pipeline. The absolute flux density scale was set according to \citet{2012MNRAS.423L..30S}. Following calibration, images for each sub-band were generated individually within \texttt{SPAM} with direction-dependent calibration applied. Subsequently, the six calibrated sub-bands for each frequency band were jointly imaged using the \texttt{wsclean} package \citep{offringa_2014MNRAS.444..606O}, producing final images at multiple angular resolutions (see Table~\ref{tab:radio_imaging_parameter} and Fig.~\ref{all_band_radio}). All radio images were corrected for the primary beam response. CASA\footnote{\url{https://casa.nrao.edu/}} viewer was used to measure the properties of the radio sources. The flux densities (S) were measured within the 3$\sigma_{\rm rms}$ contours and $\sigma_S = \sqrt{\sigma_c^2 + N_{\rm beam}\sigma_{\rm rms}^2}$ were used to calculate the error in the flux density measurements, where $\sigma_c = 0.1S$ represents error due to calibration uncertainties and $N_{\rm beam}$ is the number of beams across the diffuse emission.

\begin{table*}
\caption{Parameters of the radio images.} 
\label{tab:radio_imaging_parameter}  
\centering
\begin{tabular}{lccccccr} 
\hline
Frequency band & Central frequency (MHz) & Weighing & Robust & UV-taper (arcsec) & Synthesized beam & rms ($\mu$Jy\,beam$^{-1}$)\\
\hline
Band 3 & 400 & briggs & 0 & 10 & $13.36'' \times 10.88''$; PA: 34.97$^\circ$ & 80 \\
Band 4 & 700 & briggs & -2 & 5 & $9.11'' \times 4.84'' $; PA: 3.54$^\circ$ & 130 \\
Band 5 & 1260 & briggs & 0 & -- & $2.99'' \times 2.56'' $; PA: -53.58$^\circ$ & 41 \\
\hline
\vspace{-0.3cm}
\end{tabular}
\end{table*}

\begin{figure*}
 \includegraphics[width=0.45\textwidth]{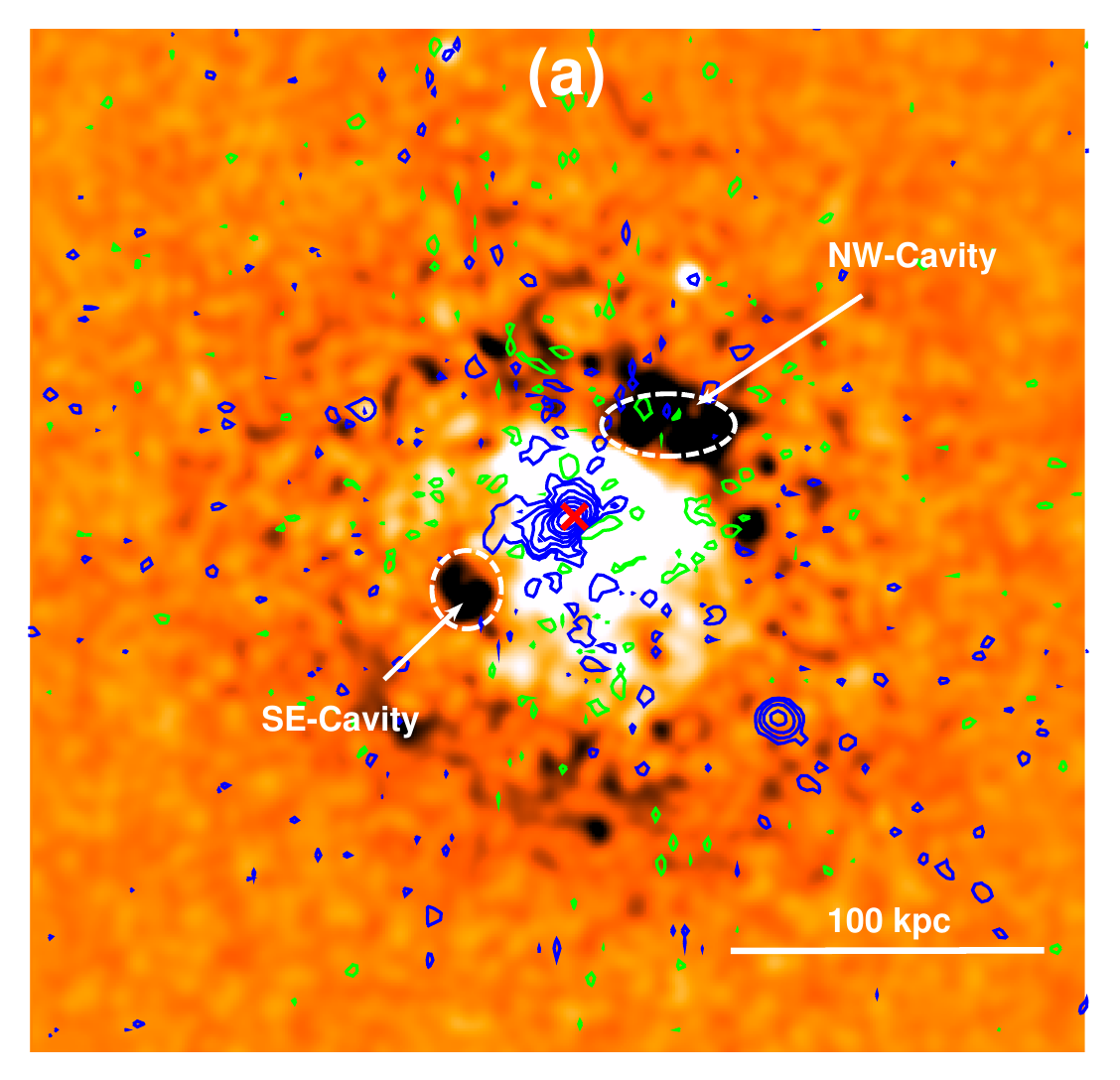}
  \includegraphics[width=0.45\textwidth]{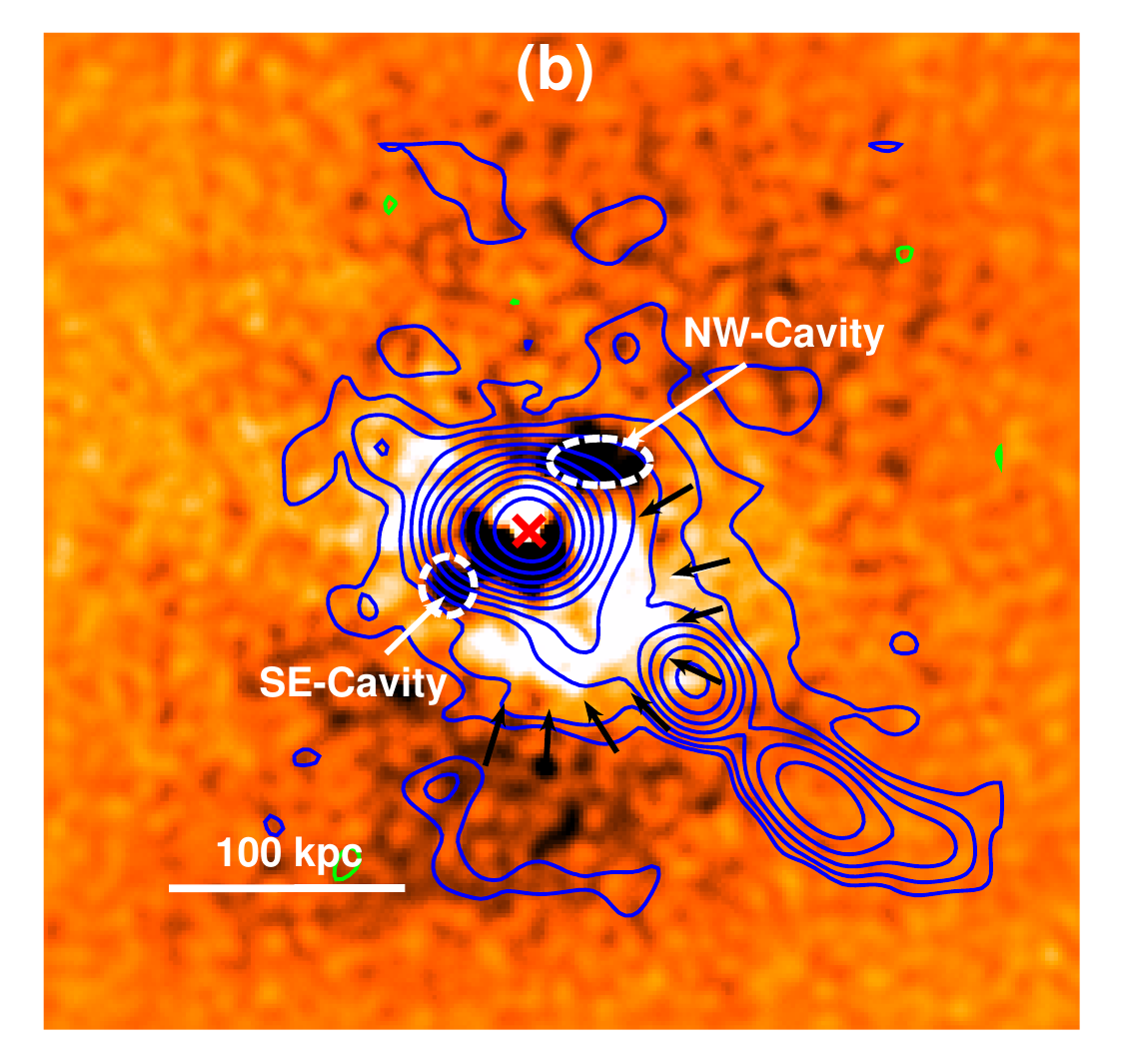}
  \includegraphics[width=0.40\textwidth]{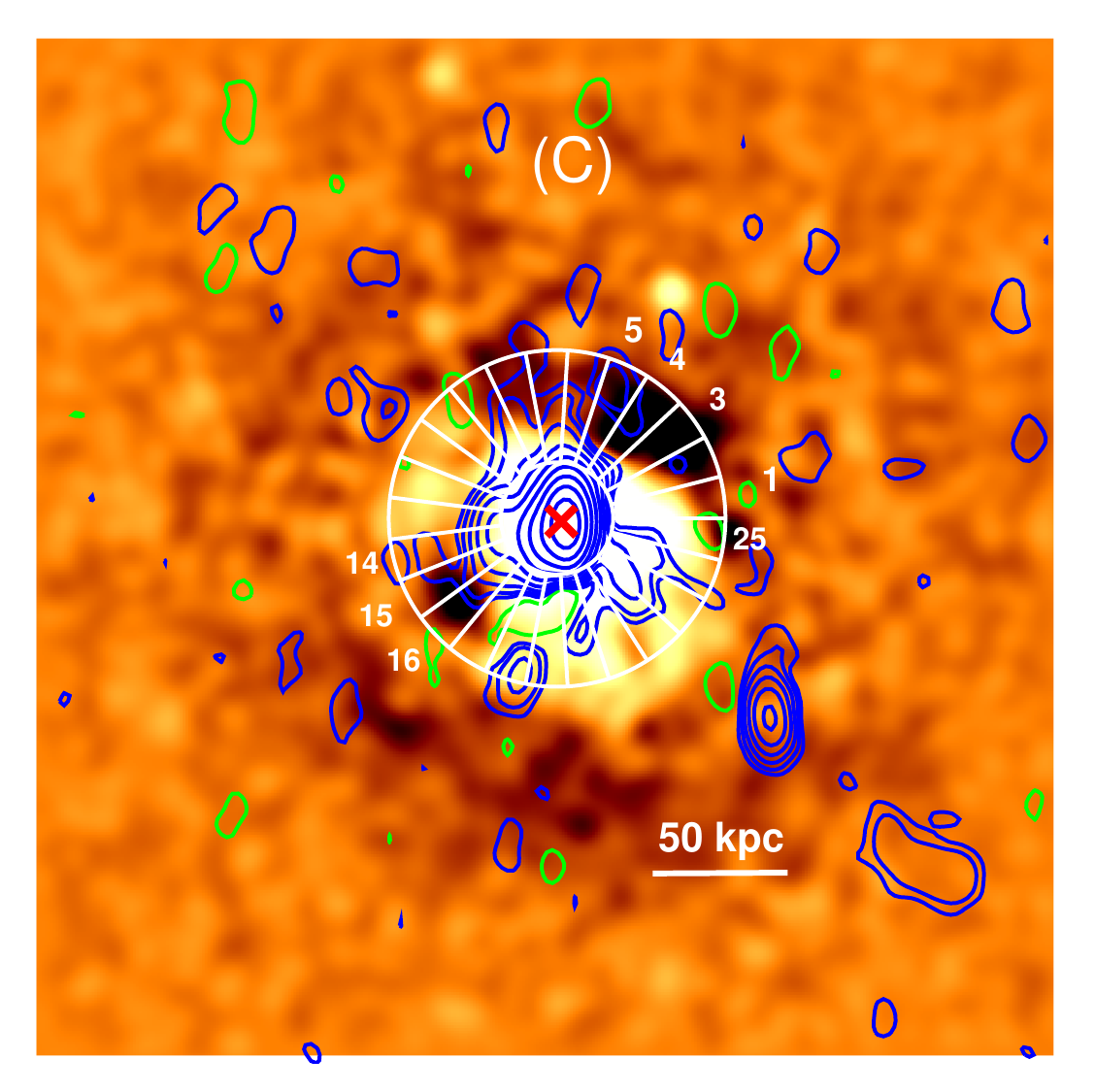}
  \includegraphics[width=0.50\textwidth]{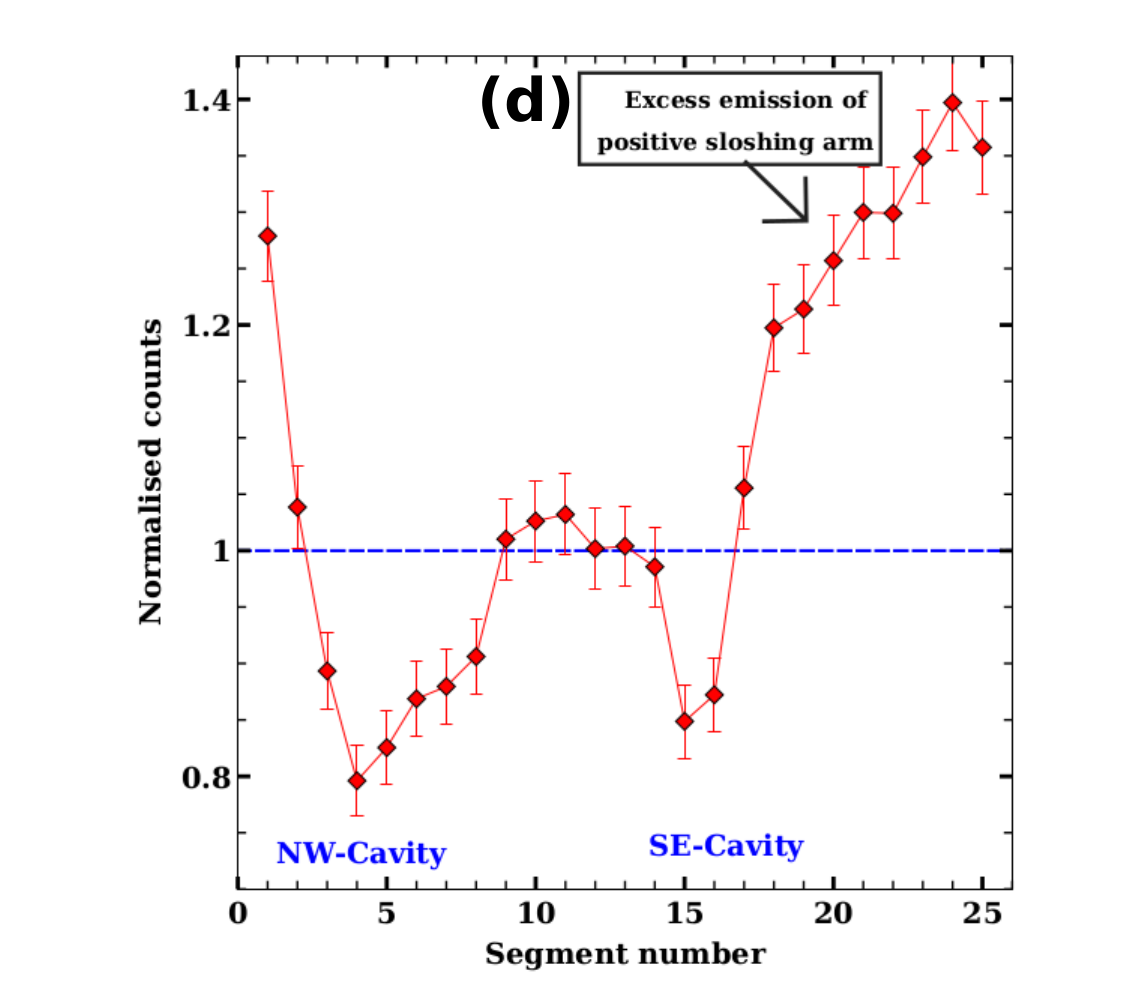}
   \caption[Residual images] {\textit{(a)}: An unsharp-mask image of RXCJ1558 clearly shows NW and SE cavities. The image is overlaid with blue uGMRT Band 5 radio contours plotted at levels of $3\sigma \times (1, 3, 9, 27, \ldots)$ and negative $-3\sigma$ contours shown in green with $\sigma = 41~\mu$Jy beam$^{-1}$. \textit{(b)}: 3.8$\arcmin$ $\times$ 3.8$\arcmin$ cropped 0.5-3 keV smoothed residual image of RXCJ1558 derived by subtracting the $\beta$-model, clearly discloses the clockwise sloshing structure in the ICM (black arrows), starting from cluster centre. This image is overlaid with blue contours of Band 3 plotted at $3\sigma \times (1, 2, 4, 8, \ldots)$, with $\sigma = 80~\mu$Jy\,beam$^{-1}$, and green contours shows negative $3\sigma$ level. \textit{(c)}: An unsharp mask image overlaid with different segments used to study the variation in counts at the cavity locations overlaid with Band 4 blue contours plotted at $\sigma \times (3, 6, 12, 24, \ldots)$. \textit{(d)}: A plot of the normalised counts versus sector number. The horizontal dashed blue line shows the average normalised counts without deviations.}
    \label{new_resid}
\end{figure*}

\section{Results}
\label{results}
\subsection{X-ray imaging analysis}
\label{sect3.1}
To identify and confirm different features in the RXCJ1558 cluster, we created the exposure-corrected, background-subtracted 0.5-3.0 keV X-ray image. This was achieved by subtracting the background image from the X-ray count image and then dividing it by the exposure map. The resultant image (Fig.~\ref{raw}, left panel) indicates two X-ray depressions highlighted by arrows along with a luminous X-ray point source coinciding with the location of the BCG, PKS 1555-140, as seen in the Pan-STARRS DR1 \textit{r}-band optical image (Fig.~\ref{raw}, right panel). Within the same image (Fig.~\ref{raw}, left panel) at the centre (40\arcsec$\times$ 40\arcsec), we identify notable X-ray substructures (see inset) that differ from the anticipated smooth distribution of the X-ray surface brightness. These are indicated by green arrows, including the location of the BCG. Such substructures often trace features like gas sloshing. The previous study done by \cite{sonkamble2017xray} identified two cavities in this cluster, while a recent study by \cite{2024ApJ...961..134U} identified a single cavity. To check whether these cavities are artifacts or X-ray depressions within ICM due to the AGN outburst and explore any hidden features within the cluster, we employed the four methods such as unsharp mask image \citep{2009ApJ...705..624D}, residual image \citep{2010ApJ...712..883D}, \texttt{CADET} \citep{2024MNRAS.527.3315P}, and count rate analysis \citep{2021MNRAS.504.1644P}.

An unsharp mask image was created by subtracting a 30$\sigma$ Gaussian kernel smoothed X-ray image from that of the image smoothed by a 3$\sigma$ Gaussian kernel. The resulting unsharp-masked image, shown in Fig.~\ref{new_resid}(a), clearly reveals two X-ray surface brightness depressions, one to the north-west (NW) and another to the south-east (SE) of the central X-ray peak. Furthermore, Band 5  uGMRT radio contours, appears as a point-like emissions with slight extension along the SE cavity, are overlaid. These features are consistent with those reported by \cite{sonkamble2017xray}. We determined the position of the earlier identified SE cavity at distances of $\sim$36 kpc and the newly detected NW cavity at $\sim$42 kpc, along with a sloshing arm.
 
To check whether these depressions in the unsharp image are indeed cavities rather than image processing artifacts, we adopted an alternative approach by utilising the \texttt{Sherpa} fitting function available in CIAO to create a residual image by subtracting a smooth 2D $\beta$-model from the 0.5–3.0 keV background-subtracted image. During image fitting, parameters such as ellipticity, position angle, normalisation angle, and background were kept free. Fig.~\ref{new_resid}(b) is the resultant residual image that confirms the SE and NW cavities and their locations, along with the sloshing arm extending toward the SW direction, as seen in the unsharp mask image. On the same image, uGMRT Band 3 contours are overlaid in blue colour showing an orientation from north-east to south-west.

The cavities appear to be embedded within the wider radio diffuse emission of Band 3. The distribution and properties of this radio emission are described in Sections~\ref{radio} and \ref{disAGN}. 
These features were further examined using the newly developed Cavity Detection Tool \citep[\texttt{CADET};][]{2024MNRAS.527.3315P} for the automated identification of X-ray cavities, as detailed in Appendix~\ref{cadet}. \texttt{CADET} combines unsharp masking, wavelet decomposition, and machine learning techniques to robustly identify surface brightness depressions in X-ray images. With this method, we confirmed the presence of two cavities at locations consistent with those identified using the techniques described above. All lines of evidence support the presence of a second cavity (NW cavity) within this cluster and highlight the consistency between the automated and manual approaches, reinforcing the reliability of the analysis.

Further, to find the significance of these cavities, we performed a count rate statistical analysis followed by the techniques used in \citet{10.1111/j.1365-2966.2011.20358.x,2019MNRAS.485.1981V,2024JApA...45...23S}. 
In this method, we extracted X-ray counts from an annular ring ranging from 10\arcsec to 30\arcsec in radius, divided into 25 equal-area sectors, and are shown in Fig.~\ref{new_resid}(c). The Figure also shows the uGMRT Band 4 contours overlaid in blue colour.
The normalised counts for each sector were plotted as a function of the sector number (Fig.~\ref{new_resid} d), delineating two surface brightness depressions between sectors 3 to 5 and sectors 14 to 16. The blue dashed horizontal line in the plot indicates the anticipated normalised counts assuming no such deviations. The excess emission between sectors 20 to 25 is likely due to the sloshing arm. The surface-brightness decrement at each cavity (D) was computed using the expression used by \cite{2021AN....342.1207U}
\begin{equation}
              D = (1 - \frac{S_{c}}{S_{s}} ) \times 100 \%
\end{equation}   
where $S_{c}$ is the surface-brightness measured in the sectors covering the cavity and $S_{s}$ is the surface-brightness measured in the immediate surroundings. The decrements are $D_{3,4,5}$ (NW cavity) $\sim$15$\%$ and $D_{14,15,16}$ (SE cavity) $\sim$14.2$\%$.
 
Further, we calculate the signal-to-noise (S/N) for each cavity using the expression of \cite{2024ApJ...961..134U}
\begin{equation}
     S/N = \frac{{|C_{c} - C_{s}|}}{{\sqrt{C_{c} + C_{s}}}}
\end{equation}
where, $C_{c}$ is the number of net counts within the cavity region, $C_{s}$ is the average number of counts in similar regions with the same size and at the same projected distance from the X-ray peak.
For NW cavity we measure $S/N = 6.6$ (with $C_{c}=1153$ and $C_{s}=1450$) and for SE cavity we get $S/N = 5.03$ (with $C_{c}=903$, $C_{s}= 1237$). Our calculated S/N for SE cavity is in good agreement with the value reported by \cite{2024ApJ...961..134U}.

\begin{figure*}
\includegraphics[scale=0.35]{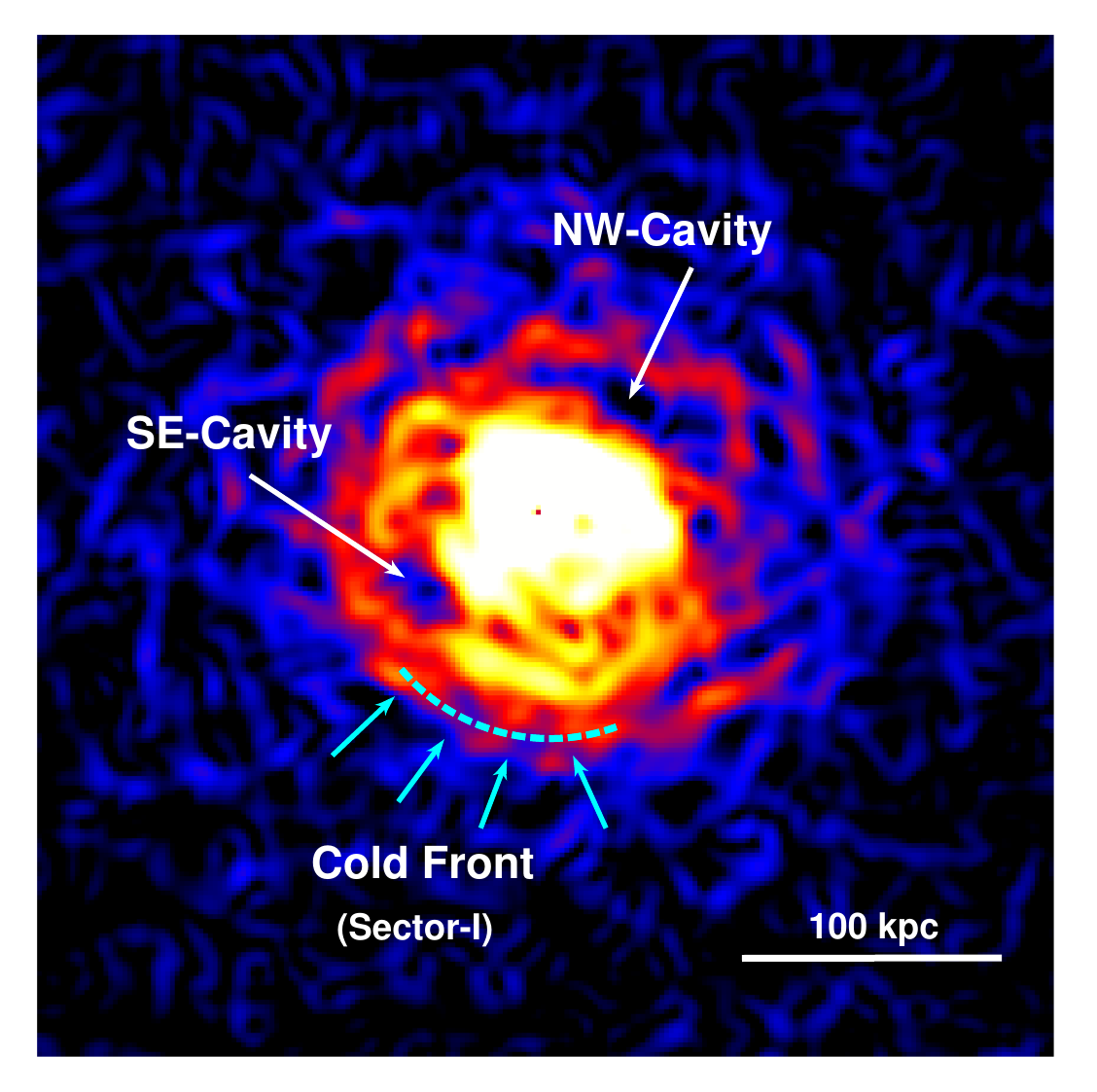}
\includegraphics[scale=0.45]{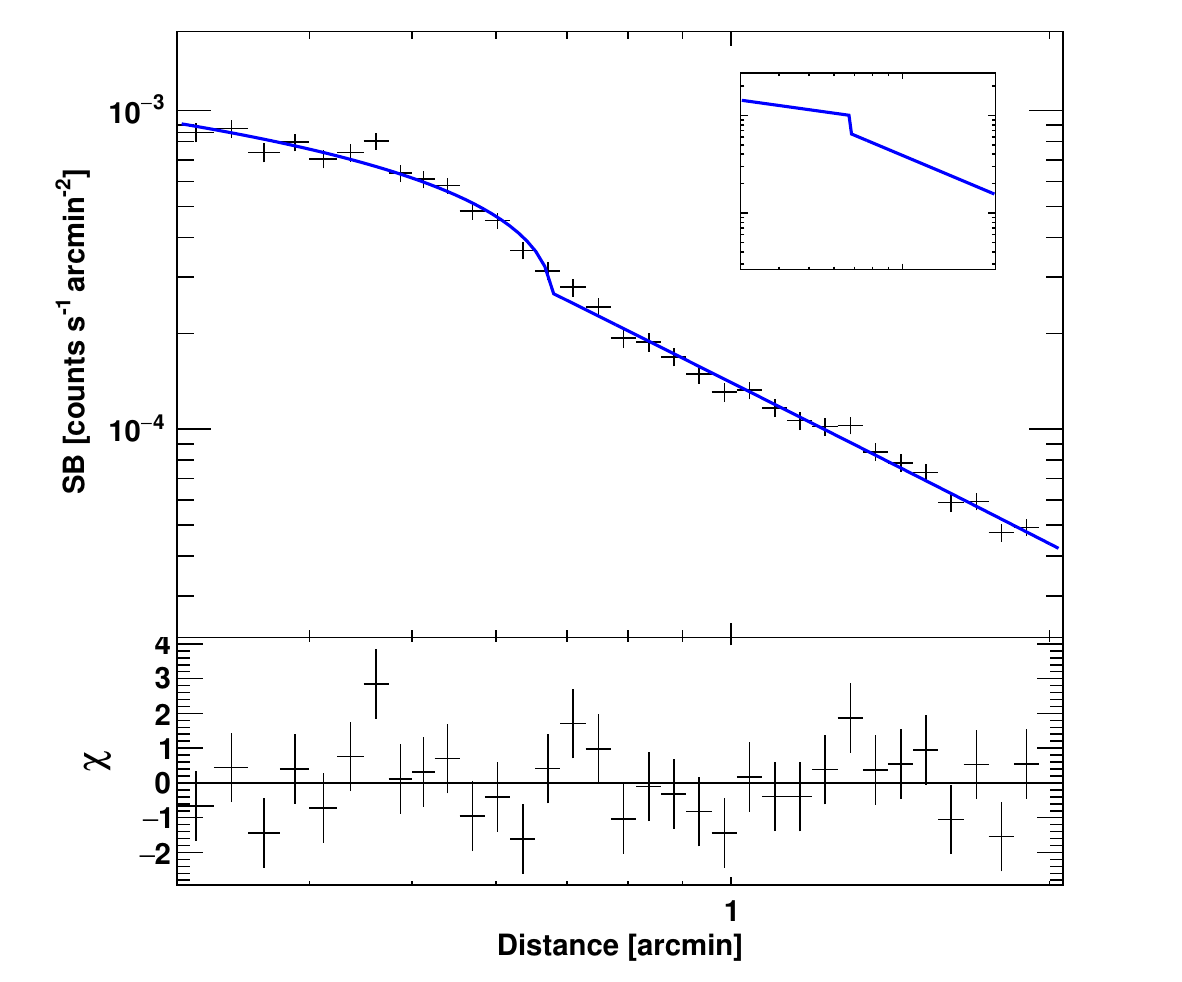}
\includegraphics[scale=0.45]{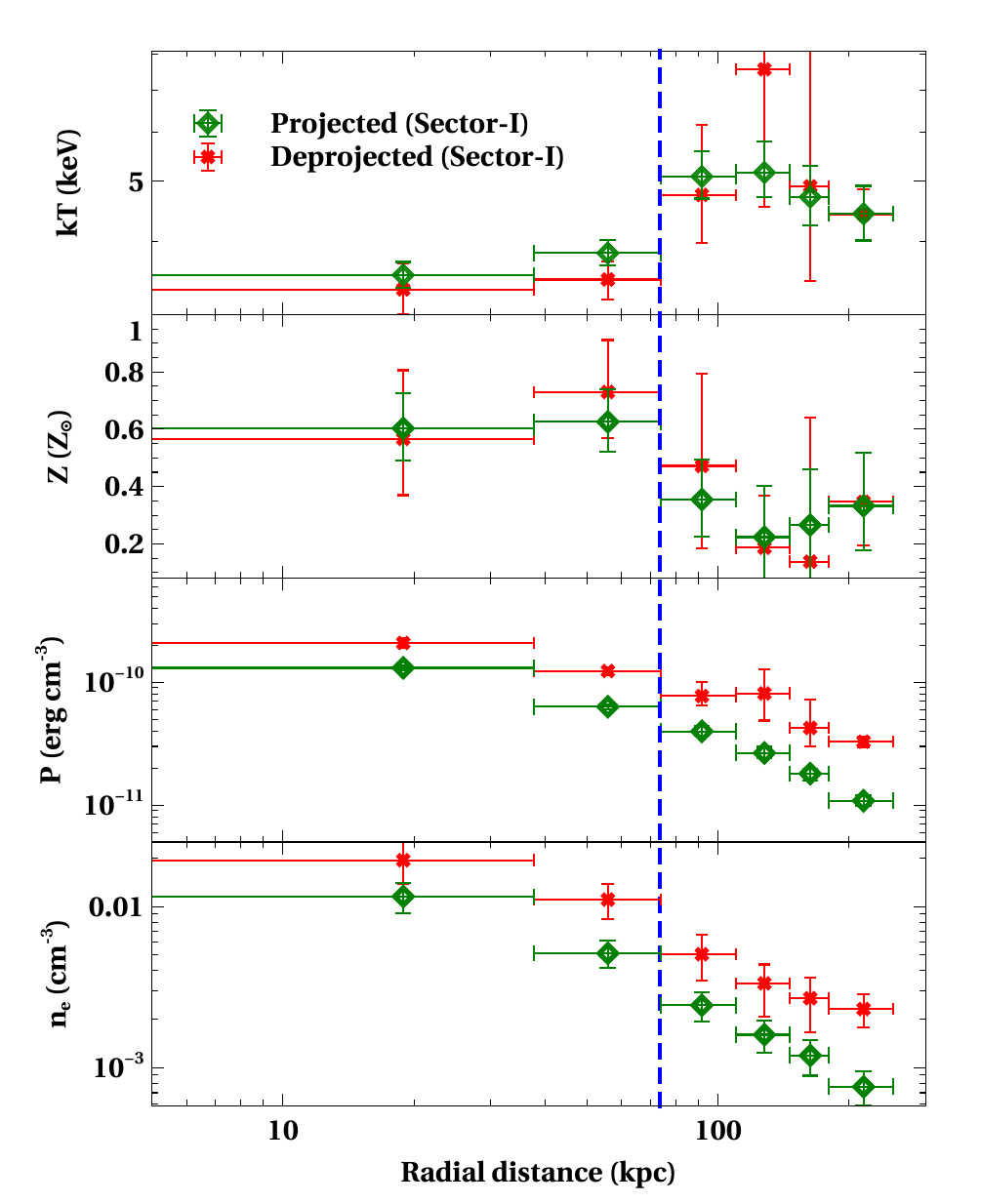}
\caption{\textit{Upper left panel} : A GGM filtered image of RXCJ1558 on a scale of 3$\sigma$. A cold front is apparent across Sector-I, marked by a cyan arc along with NW and SE cavities. \textit {Upper right panel} : X-ray surface brightness profiles in the 0.5–3.0 keV energy band extracted along Sector-I. This profile was fitted with broken-power law density model shown in solid blue line. The insets in the panel show the corresponding 3D simulated gas density model. \textit{Lower panel} : Thermodynamical profiles of temperature, metallicity, pressure, and electron density plotted as a function of radial distance. Green open diamonds and red crosses represent projected and deprojected profiles. The vertical blue dashed line indicates the location of the cold front.}
\label{cf}
\end{figure*}

\subsection{X-ray surface brightness profile} 
{To check the presence of any surface brightness edge associated with the sloshing arm, we have generated a Gaussian Gradient Magnitude (GGM) image which was created with a smoothed Gaussian kernel of 3$\sigma$ pixel from the exposure-corrected, background-subtracted X-ray image, displayed in Fig.~\ref{cf} (upper left). This figure clearly reveals an edge along the Sector-I (218$^{\circ}$ to 280$^{\circ}$) at a distance of $\sim$72 kpc from the BCG.

 The edge obtained above was verified using the broken-power law density model using \texttt{PROFFIT-v1.4} \citep{2011A&A...526A..79E} package. We extracted the surface brightness profile of the X-ray emission from Sector-I and the resultant profile is shown in Fig.~\ref{cf} (upper right), which indicates a sharp edge. This was then fitted with a single broken power-law density model parametrised as

\begin{table*}
\caption{Best fitting parameters of broken-power law model along Sector-I. }
\label{Table_cf}   
\begin{tabular}{cccccc}
\hline
$\alpha1$ & $\alpha2$ & r$_{cf}$ (arcmin) & Norm $(10^{-3})$ & Compression ($C$) & $\chi^2/{\rm dof}$
 \\
\hline
0.44$\pm$ 0.09 & 1.3$\pm$ 0.2 & 0.67$\pm$ 0.06 & 84.0$\pm$ 0.3 & 1.52$\pm$ 0.09 & 33/ 27 \\
\hline
\end{tabular}
\end{table*}

\begin{equation} 
n (r) = 
 \begin{cases}
     Cn_{0} {(\frac{r}{r_{cf}})}^{-\alpha1} &, \, \text{if} \,\,  {r < r_{cf}} \\
     n_{0} {(\frac{r}{r_{cf}})}^{-\alpha2}  & , \, \text{if} \,\, {r > r_{cf}}
     \end{cases}
     \label{SBequation}
\end{equation}

where, $n (r)$, the electron number density at the projected distance $r$, $n_{0}$ the density normalization, $C$ the density compression factor of the surface brightness edge , while $\alpha1$ and $\alpha2$ being the power-law indices, and $r_{cf}$ the radius corresponding to the surface brightness edge . All parameters during the fit were allowed to vary. The best-fitted broken power-law model yields a reduced $\chi^2$/d.o.f = 1.22. This model clearly indicates the presence of a surface brightness discontinuity at a radius of $\sim$72.4 kpc from the X-ray peak. The discontinuity is arc-shaped, which coincides with the observed edge of radio diffuse emission and spatially coincides with a metallicity drop. We verified that the location of the discontinuity ($r_{cf}$) and the associated jump parameters remain consistent even when increasing the arc sector size in the same direction, confirming the robustness of the fit. The best-fitted broken power-law model is shown in Fig.~\ref{cf} (upper right), and the corresponding parameters, along with fitting errors at 68\% confidence, are listed in Table~\ref{Table_cf}. This analysis supports the presence of the surface brightness edge at $\sim$72.4 kpc, likely associated with sloshing motions in the ICM.}

\subsection{Spectral analysis}

{To determine the nature of the surface brightness edge detected above, we derived the projected thermodynamical quantities, including temperature (kT), metallicity (Z), pressure ($p$), and electron density ($n_{e}$) along the sector-I as shown in Fig.~\ref{cf} (lower panel)}. To achieve this, we used \texttt{specextract}\footnote{\url{https://cxc.cfa.harvard.edu/ciao/ahelp/specextract.html/}} script in CIAO to extract spectra from this sector-I in the energy range of 0.5-7.0 keV from seven wedge-shaped regions centred on the BCG. The background spectrum was obtained from a blank sky background file. Response matrices (RMF) and photon-weighted effective area (ARF) files were generated using the same script. Each annulus spectrum was then fitted independently with a single-temperature collisionally ionized diffuse gas model (ATOMDB code) \texttt{APEC} \citep{2001ApJ...556L..91S} with the photoelectric absorption model \texttt{TBABS} \citep{2000ApJ...542..914W}. During the fitting, we allowed temperature (keV), metallicity (\Zsun), and normalisation (cm$^{-5}$) to vary while keeping hydrogen column density and redshift at 1.05 $\times 10^{21}$ cm$^{-2}$ \citep{2016A&A...594A.116H} and 0.097, respectively.
  
Then we measured the electron density $n_{e}$ (cm$^{-3}$) using the normalisation obtained from the \texttt{APEC} and the expression \citep{2022ApJ...938...51A} 

\begin{equation}
    n_e  = \biggl[1.2 \ N \times 4.07\times10^{-10}(1+z)^2 \\ \biggl (\frac{D_A}{\rm{Mpc}}\biggl)^2\biggl(\frac{V}{\rm{Mpc^3}}\biggl)^{-1}\biggl]^{1/2} 
\label{eq:density}
\end{equation}

Here, $D_{A}$ represents the angular diameter distance of the source and $n_{e}$ and $n_{H}$ are the electron and hydrogen densities, respectively. We assume $n_{e} / n_{H}$ $\sim$ 1.8. After estimating the values of the temperature and electron densities corresponding to each annulus, we compute the pressure using $p = nkT$. Here, $n$ is the total gas density ($n \sim$ 1.92$n_{e}$). The resultant projected profiles are shown in Fig.~\ref{cf} (lower panel) with green open diamonds. A temperature jump has been confirmed at $\sim$72.4 kpc from $3.83\pm0.17$ keV to $5.09\pm0.39$ keV, while pressure was found to remain continuous across this region. As a result, this surface brightness edge is characterised as a cold front. 

\begin{figure*}
\centering
\includegraphics[scale=0.80]{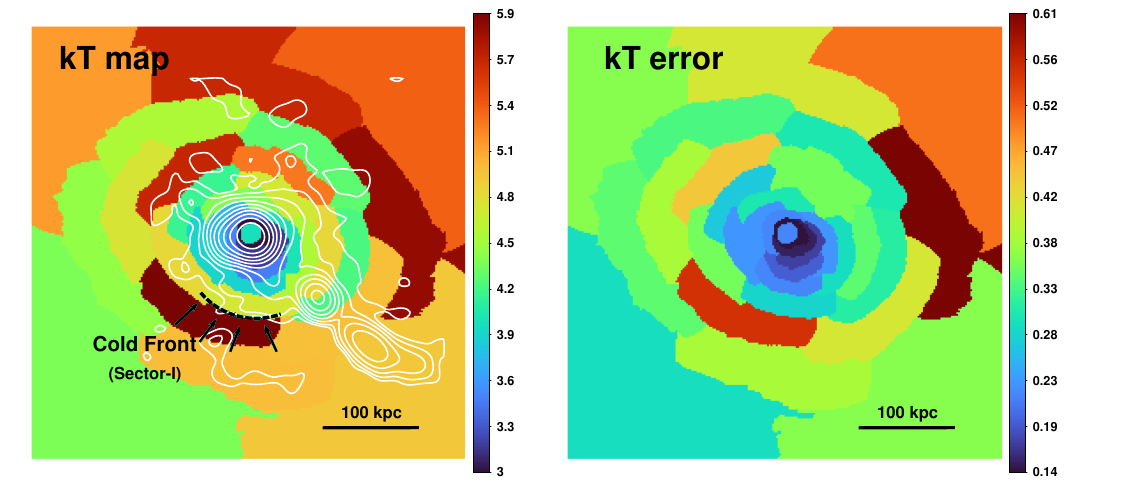}
\includegraphics[scale=0.80]{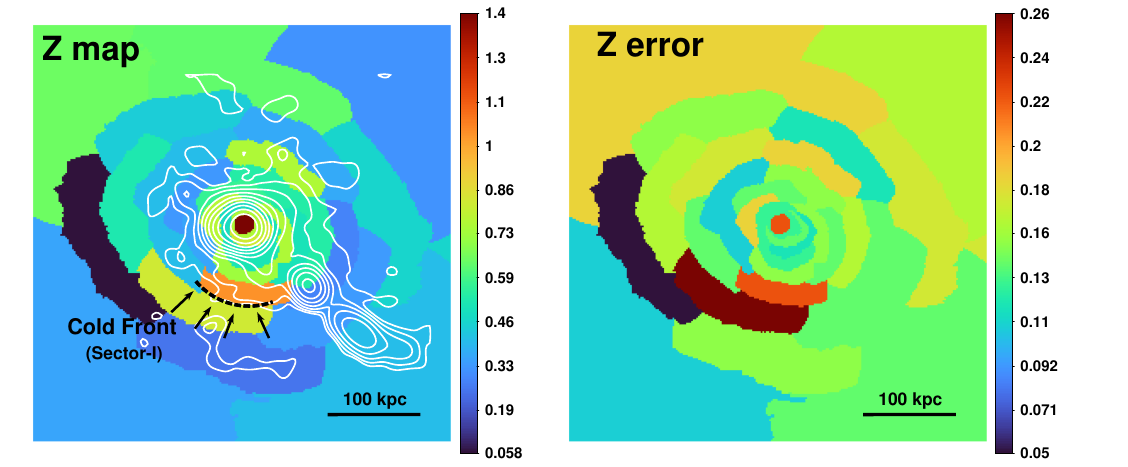}
\caption[kT and Z Map]{\textit{Upper panel:} 2D temperature map (upper left), its error (upper right) of the ICM obtained from contour binning technique. \textit{Lower panel:} 2D metallicity map (lower left) and its error (lower right). Colour bar in temperature and its error show in unit of keV. Colour bar in metallicity and its error show in the unit of Z$_{\odot}$. On both maps, uGMRT Band 3 radio contours are overlaid, and the cold front is marked by black dotted arc.}
\label{added_map}
\end{figure*}

To verify that the observed temperature discontinuity is not due to the projection effects, we employed a deprojection technique to account for contamination from overlying ICM layers. For the previously extracted spectra from seven wedge-shaped regions, we incorporated the \texttt{PROJCT} model \citep{2003ApJ...585..227B} into a simultaneous spectral fitting routine. This allowed us to derive deprojected values for temperature, metallicity, and normalisation. Using these parameters, we further computed the deprojected electron density and pressure profiles, analogous to the projected profiles discussed earlier. The resulting thermodynamic profiles are shown in Fig.~\ref{cf} (lower panel) with red crosses. A clear temperature jump is evident at the location of the cold front from 3.47$\pm$0.24 keV to 4.78$\pm$0.70 keV. We also notice a drop in the metallicity from 0.63$\pm$0.10 to 0.35$\pm$0.13 $\Zsun$ and a drop in electron density from 11.0$\pm$2.8 to 5.0$\pm$1.5 (10$^{-3}$ cm$^{-3}$), while the pressure remains continuous across the discontinuity, indicating the presence of a cold front. Thus, both of these methods confirm that the ICM in the centre of RXCJ1558 is disturbed, and a sloshing motion creates a cold front. We also compare the density compression factor ($C$) derived from the surface brightness ($S_X \propto n_e^{2}$) with the density jump obtained from deprojected spectral analysis. The spectral analysis yields $C_{\rm{spec}}$ = $(11 \pm 2.8) \times 10^{-3} / (5 \pm 1.5) \times 10^{-3} = 2.2$, which is about 44\% higher than the compression factor from the surface brightness ($C_{\rm{sb}} = 1.52$), indicating a stronger density discontinuity when measured from the spectral analysis.

\begin{figure*}
    \includegraphics[scale=0.24]{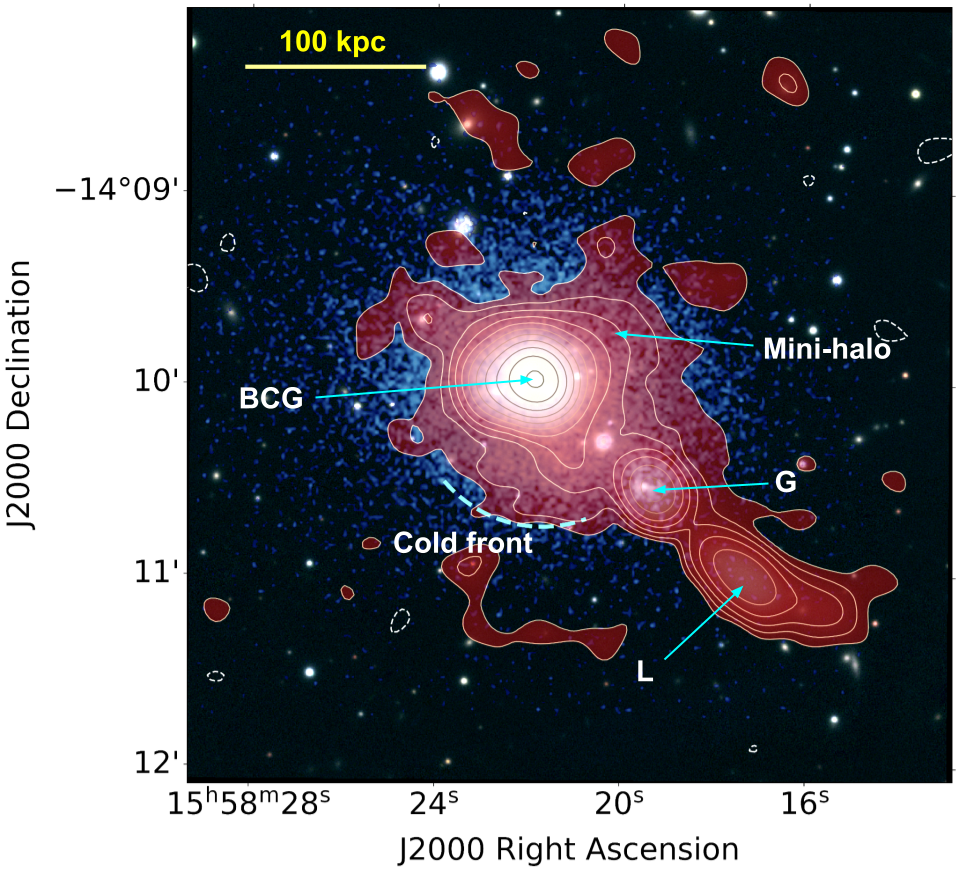}
    \includegraphics[scale=0.32]{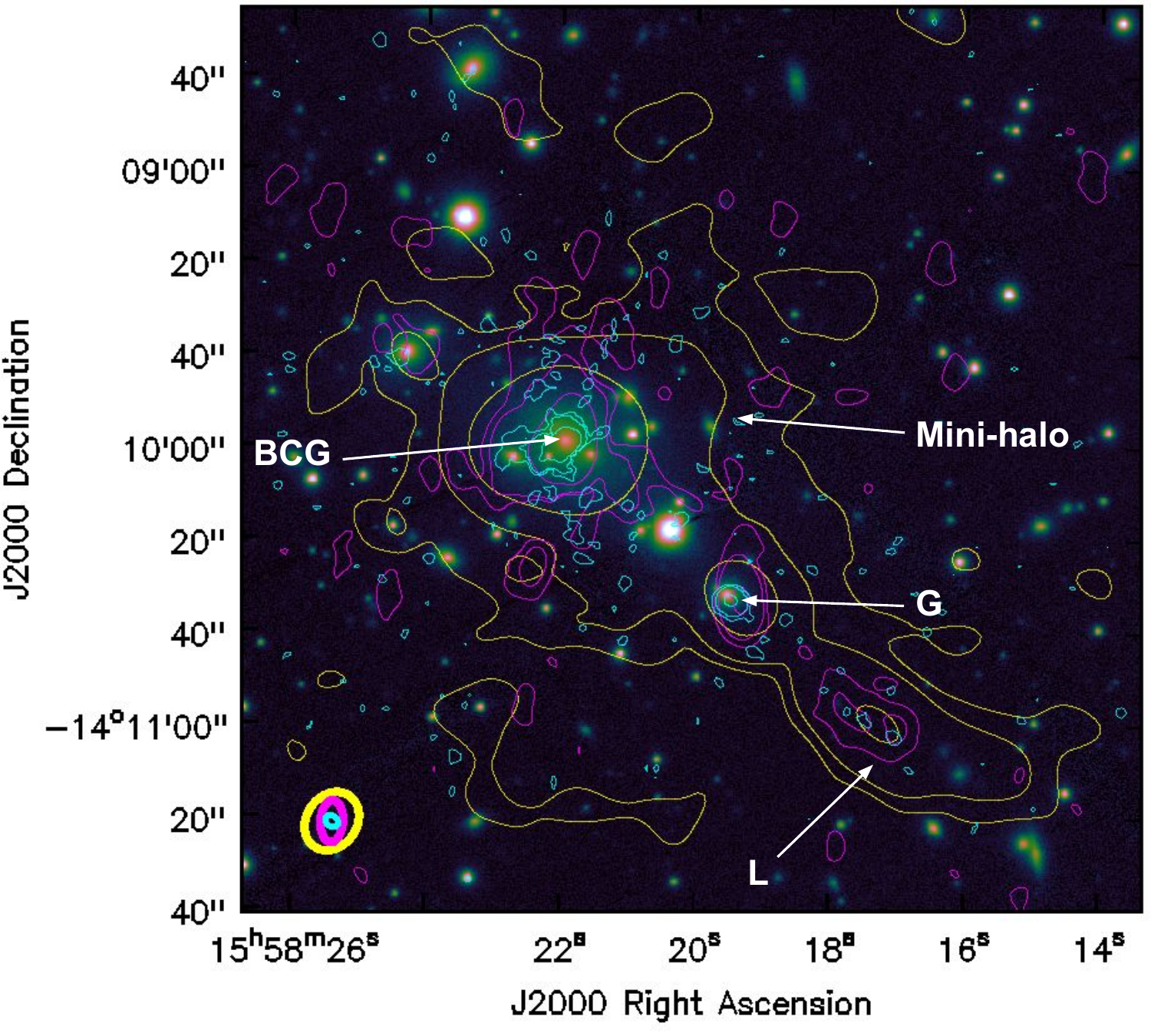}
    \caption{ \textit{Left panel}: Colour composite of the RXCJ1558 field with Pan-STARRS DR1 \textit{i}, \textit{r}, \textit{g} bands in the background red and blue rasters show low resolution uGMRT Band~3 radio and \chandra\, X-ray images, respectively. Contours from the same radio map are plotted at $3\sigma \times (1, 2, 4, 8, \ldots)$, with $\sigma = 80~\mu$Jy\,beam$^{-1}$. $-3\sigma$ dashed contours are plotted. \textit{Right panel}: A Pan-STARRS DR1 \textit{i}-band image of the RXCJ1558 field overlaid with uGMRT radio contours at $3, 9$ and 81$\sigma$ for Bands~3 (yellow), 4 (magenta), and 5 (cyan), with $\sigma = 80$, $130$, and $41~\mu$Jy\,beam$^{-1}$, respectively.} 
    \label{fig:R1558_composite}
\end{figure*}

\subsection{Temperature and metallicity map}

To examine the 2D spatial variations in the temperature and metallicity distribution of the ICM in the same region as shown in GGM image above, we derived temperature and metallicity maps. These maps were created using the contour binning algorithm (\texttt{CONTBIN}) developed by \cite{2006MNRAS.371..829S}. We first excluded the point sources from the image. The technique then identifies the brightest pixels in the X-ray image and grouped into small initial bins containing pixels of the same surface brightness reached until it reaches the S/N up to 50 (2500 net counts in each bin). Here we get the total 33 number of bins in the 0.5-3.0 keV image. We generated the spectra for each region using the same method mentioned above. The derived temperature and metallicity, along with their respective uncertainties are shown in Fig.~\ref{added_map}. The statistical errors in the central/peripheral regions in the temperature map are 6\%/14\%, respectively. The statistical errors for the metallicity map are 22\%/30\% for the central/peripheral regions.

The temperature distribution demonstrates that the innermost ICM is the coolest, measuring 3.5 keV, with a gradual increase in temperature moving outward, and it reaches up to $\sim$6.6 keV. The temperature map shows that RXCJ1558 is a cool-core cluster, with a core - outskirts temperature difference of about 3 keV (i.e factor of 2), indicative of its cool-core nature \citep{2010A&A...513A..37H}. The map also indicates a notable temperature increase toward the south, with temperatures rising sharply from 4.8$\pm$0.4 keV to 5.8$\pm$0.5 keV similar to what was seen in our radial thermodynamic profile attributing the cold front. The spiral arm extends from the cool core into the ICM surrounding it. The map clearly reveals the sloshing arm, with temperatures in the range of $\sim$3.5-4 keV. We observe that the cooler temperature region extends toward the south-east, following the spiral arm. Notably, these spiral patterns occur in regions of relatively low temperature, consistent with results reported for other clusters exhibiting similar spirals \citep{2017ApJ...837...34U, 2019ApJ...871..207U}. The uGMRT Band 3 mini-halo radio contours are overlaid on the 2D temperature and metallicity maps, revealing a clear link between the thermal and non-thermal emissions. In particular, the alignment between the radio structures and the temperature and the X-ray surface brightness (see Fig.~\ref{fig:R1558_composite}, left panel) indicates that both are likely driven by processes occurring within the ICM. Furthermore, the mini-halo radio emission appears to be confined within the cold front, suggesting a close link between the observed features. A more detailed interpretation of these structures, as well as the connection between the thermal and non-thermal components, are presented and discussed in the following sections. The metallicity map illustrates the complex distribution resulting from the sloshing motion of the ICM. In the map, the sloshing arm shows a high abundance of metals, quantified as 1.1$\pm$0.25 Z$_{\odot}$, in contrast to the surrounding ICM.

\subsection{Large-scale radio emission}
\label{radio}

Fig.~\ref{fig:R1558_composite} (left panel) presents a colour composite image of RXCJ1558. The background comprises Pan-STARRS DR1 optical images in the \textit{i}, \textit{r}, and \textit{g} bands. Overlaid are the \chandra\, X-ray image (blue) and the low resolution uGMRT Band~3 radio image (red), along with the contours of same radio map (see Table~\ref{tab:radio_imaging_parameter} for imaging details). The low resolution radio map clearly reveals diffuse emission surrounding the central BCG, indicative of a possible radio mini-halo. A candidate mini-halo or large scale fossil lobes in this system has previously been suggested based on TGSS survey image \citep{2015MNRAS.453.1201H}. However, the limited angular resolution and sensitivity of the survey data hindered the identification to the full extent and substructure of the diffuse emission, and thus, the mini-halo nature could not be confirmed in their analysis. In contrast, the present uGMRT observations reveal a well defined, extended diffuse structure with a largest linear size (LLS) of approximately $113''$ ($\sim$200 kpc). The peak of this diffuse radio emission coincides spatially with both the central BCG and the X-ray emission peak, strengthening the mini-halo interpretation.

The radio emission in this cluster is found to be elongated towards the south-west. The extended part may have been produced due to overlapping radio contours from two other radio sources (G and L). These sources are located at projected distances of $\sim$95 kpc and $\sim$170 kpc and are seen towards the south-west of the central radio peak. The first source (G) coincides with an optical counterpart identified as a galaxy (WISEA J155819.53-141031.6 at redshift of 0.1122) in the NED\footnote{\href{https://ned.ipac.caltech.edu/}{NASA/IPAC Extragalactic Database (NED)}}. The more distant patch (L) lacks any optical counterpart in either NED or the Pan-STARRS DR1 optical image (see Fig.~\ref{fig:R1558_composite}, right panel). This source exhibits a LLS of $\sim$105~kpc and an ultra-steep spectral index ($\alpha_{400MHz}^{700MHz}$) of $-1.78 \pm 0.41$. Consequently, it remains unclassified in this study. However, it is plausible that this patch represents an aged lobe originating from galaxy~G, with its potential counter-lobe now blended with the diffuse emission associated with the mini-halo. In this case, the Band~3 mini-halo flux density and the corresponding 1.4~GHz radio power should be regarded as upper limits. As an alternative interpretation, the radio emission associated with galaxy~G may instead arise from a narrow-angle tailed (NAT) radio galaxy. Based on the redshift difference, galaxy~G is unlikely to be a cluster member; nevertheless, a NAT morphology cannot be ruled out, although it would most plausibly correspond to a background system projected along the line of sight.

The extent of the radio emission associated with the BCG is around 25 kpc (Fig.~\ref{new_resid} a) and exhibits flux densities of $363.5 \pm 36.4$~mJy, $255.9 \pm 25.6$~mJy, and $283.3 \pm 28.3$~mJy at uGMRT Bands~3, 4, and 5, respectively, as measured from our high-resolution maps at 3$\sigma$. According to \cite{2015MNRAS.453.1201H}, this source is a Gigahertz-Peaked Spectrum source that also exhibits variability. The most distant radio patch (L) observed to the south-west shows a flux density of $15.4 \pm 1.6$~mJy at Band 3 and $5.7 \pm 0.7$~mJy at Band 4, and remains undetected at Band~5. To isolate the emission of the candidate mini-halo, we subtracted the flux density of the central BCG (measured from high-resolution maps) from the total flux density of the diffuse emission measured in the corresponding low resolution map. The resulting flux density of the mini-halo is estimated to be $40.9 \pm 4.1$~mJy at Band 3. The contamination from diffuse emission associated with the cavities and a possible counter-lobe of source L cannot be subtracted; therefore, the measured mini-halo flux density should be considered an upper limit. The detection of the mini-halo at Band~4 is marginal, and remains undetected in Band~5. The $K$-corrected radio power of the mini-halo at 1.4~GHz is calculated to be $(2.47 \pm 0.25) \times 10^{23}$~W\,Hz$^{-1}$ using the following relation 

\begin{equation}
    P_{\rm 1.4 GHz}=4\pi S_{\rm 1.4 GHz}D^2(1+z)^{-(\alpha+1)}
\end{equation} \label{eq: radio_power}

where $D$ is the luminosity distance and assuming a spectral index of $\alpha = -1.1$ for the radio mini-halo \citep{giacintucci_2014ApJ...781....9G, 2016MNRAS.455L..41B}. $S_{\rm 1.4 GHz}$ is the flux density at 1.4 GHz and $z$ is the redshift. The properties of the radio sources are summarised in Table~\ref{tab:diffuse_emission}.

\begin{table}
\caption{Measured properties of the different radio sources.}
\centering
\small
\begin{tabular}{lcccc}
\hline
Source & \multicolumn{3}{c}{Flux density (mJy)} & LLS (kpc)\\
 & Band 3 & Band 4 & Band 5 & Band 3 \\
\hline
BCG & $363.5\pm36.4$ & $255.9\pm25.6$ & $283.3\pm28.3$ & -- \\
Mini-halo$^{\dag}$ & $40.9\pm4.1$ & -- & -- & 205 \\
source G & $19.7\pm2.0$ & $15.9\pm1.6$ & $8.5\pm0.8$ & -- \\
source L & $15.4\pm1.6$ & $5.7\pm0.7$ & -- & 105 \\
\hline
\end{tabular}
{\raggedright
$^{\dag}$\,The flux density at Band 3 may be overestimated, as extended radio emission from the cavities related to the BCG and putative counter-lobe of source L related to galaxy G is morphologically blended within the diffuse emission of mini-halo.}
\label{tab:diffuse_emission}
\end{table}

\begin{figure*}
\centering
\includegraphics[scale=0.9]{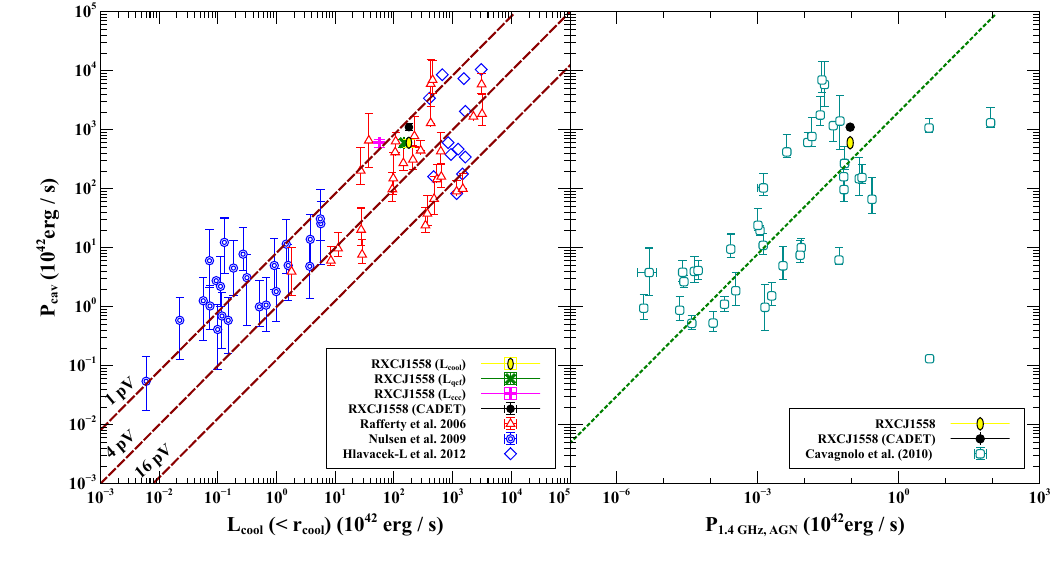}
\caption{ \textit{Left Panel:} Plot between the mechanical power ($P_{\rm cav}$) versus X-ray cooling luminosity, adopted from \citet{2012MNRAS.421.1360H} with data points from \citet{2006ApJ...652..216R,2009AIPC.1201..198N}. Additionally, we plot the positions of RXCJ1558 estimated using $R_{\rm cool}$ (yellow ellipse), $R_{\rm ccc}$ (magenta plus), and $R_{\rm qcf}$ (green cross) to explore more physically motivated estimates. \textit{ Right Panel:} Plot between the cavity power ($P_{\rm cav}$) versus 1.4 GHz AGN radio power with the sample data of \citet{2010ApJ...720.1066C}. RXCJ1558 showing best-fit balance between mechanical power and 1.4 GHz AGN radio power ($P_{\rm 1.4 GHz, AGN}$) marked by yellow ellipse (Visual) and black filled circle (\texttt{CADET}).}
\label{added_plot}
\end{figure*}

\section{DISCUSSION}
\label{disussions}
\subsection{Cavity energetics} 
\label{caveng}

The imaging analysis presented in Section~\ref{sect3.1} confirms the presence of SE and NW X-ray cavities, which are believed to have been carved out by radio jets launched from the central AGN. During this AGN outburst phase, the jets inject energy into the surrounding ICM, performing $pV$ work and inflating the cavities. These cavities then rise buoyantly through the ICM until they reach a state of pressure balance with the ambient gas. At this stage, when the buoyant rise becomes more dominant than their expansion, the cavities decouple from the jets and begin to release their stored energy into the ICM in the form of enthalpy \citep{2004ApJ...607..800B}. Therefore, study of X-ray cavities provides a unique tool to quantify the energetics of the AGN outflows. The total energy associated with the X-ray cavities can be estimated as the sum of the thermal energy of the X-ray cavities and the mechanical work done by the X-ray cavities \citep{2012AdAst2012E...6G}.

\begin{equation}
E_{\rm cav} = E_{\rm thermal} + pV = \frac{\gamma pV}{\gamma - 1}
\end{equation}

where, $p$ denotes the pressure around the X-ray cavity ($p=nkT$, with $n=1.83n_{e}$) and $V$ the volume of the cavity. Assuming the oblate or spherical shape, the volume of the cavities is calculated as V=4$\pi R_{l}^{2}R_{w}/3$ , where, $R_{l}$ is the semi-major axis and $R_{w}$ is the semi-minor axis of cavities. $R_l$ and $R_w$ were estimated using two independent approaches: (i) regular visual inspection method explained in \cite{2016ApJS..227...31S}; and (ii) an automated \texttt{CADET} based deprojection tool, which provides a more robust estimate of the three-dimensional structure of the cavities \citep{2024MNRAS.527.3315P}. Here we assume that the cavities are filled with the relativistic plasma ($\gamma$ = 4/3), which yielded the energy release of $E_{\rm cav}=4pV$. The total power injected by the central AGN into the surrounding ICM via jets or outflows was calculated by dividing the total energy content ($E_{\rm cav}$) of both cavities by their ages. The age of the cavities was calculated by three different methods described in \cite{2004ApJ...607..800B}. (i) Time ($t_{\rm sonic}$) taken by the cavities to expand with trans-sonic velocity; (ii) Time ($t_{\rm buoy}$ $\sim$ $R$ $\sqrt{SC_{D}/2gV}$) taken by the cavities to reach the current position buoyantly with terminal velocities, where $R$ is the projected distance from the centre of the cavity to the central AGN, S cross sectional area of cavity, $C_{D}=0.75$ drag coefficient, g = $\frac{2 \sigma^{2}}{R}$ gravitational potential,  where $\sigma$ is stellar dispersion velocity 280 km $s^{-1}$ taken from \cite{Pulido_2018}; and (iii) Time ($t_{\rm refill}$ $\sim$ $\sqrt{r/g}$) required to refill the displaced volume of ICM as cavities rise in the upward direction, where $r$ radius of the cavity.

Table~\ref{energetics} summarises the cavity parameters, volumes, energetics, and timescales for the NW and SE cavities in RXCJ1558, derived using two approaches. Our \texttt{CADET} derived volumes for both cavities are slightly larger than those from the visual approach, by a factor of $\sim$2.3. \cite{2024MNRAS.527.3315P} reported difference of < 0.2 dex (i.e a factor of 2) and for few objects maximum difference of  < 0.3 dex between these two methods by comparing with human made predictions by \cite{2010ApJ...712..883D}, and \cite{2016ApJS..227...31S}, showing considerable agreement. The differences in our study likely arise from the uncertainties and biases involved in visually estimating the sizes of X-ray cavities using unsharp or residual images. These differences directly impact the derived energetics and highlight the importance of accurate volume estimation in assessing AGN feedback. This difference translates into higher enthalpy estimates ($E_{\rm cav} = 4pV$), with the \texttt{CADET} method yielding $\sim$1.5 times more energy than the visual method. For instance, the NW cavity energy increases from $13.7 \times 10^{59}$ erg (Visual) to $19.2 \times 10^{59}$ erg (\texttt{CADET}).

Similarly, cavity power ($P_{\rm cav}$), calculated by dividing the enthalpy by the buoyancy timescale, shows a marked enhancement when using \texttt{CADET} based volumes. The power for the NW cavity rises from $34 \times 10^{43}$ erg s$^{-1}$ (Visual) to $73 \times 10^{43}$ erg s$^{-1}$ (\texttt{CADET}), while for the SE cavity it increases from $26.3 \times 10^{43}$ erg s$^{-1}$ to $38 \times 10^{43}$ erg s$^{-1}$. The cavity ages estimated via three different methods (sonic, buoyancy, and refill timescales) lie in the range of $\sim$4 - 15.5 $\times 10^7$ years across both cavities and methods, with \texttt{CADET} generally yielding longer refill times due to larger inferred volumes. Overall, the results from \texttt{CADET} imply a slightly more energetic AGN outburst than suggested by the visual method.

\begin{table}
    \centering
    \scriptsize	
    \caption{Cavity energetics of RXCJ1558: comparison between visual approach and \texttt{CADET} volume estimation method.}
    \begin{tabular}{c|c|c|c|c}
    \hline
    Parameters & \multicolumn{2}{c|}{Visual} & \multicolumn{2}{c|}{\texttt{CADET}} \\
    \cline{2-5}
    & NW Cavity & SE Cavity & NW Cavity & SE Cavity \\
    \hline
    R$_{l}$ $\times$ R$_{w}$ (kpc) & $20.6 \times 9.6$ & $12.1 \times 10.8$& $27.45 \times 19.08$  & $18.52 \times 16.47$ \\
    R (kpc) & 42.2 & 36.1 & 42.2 & 36.1 \\
    Cavity Vol. (10$^{68}$  cm$^{3}$) & 7.6&  3.0 & 17.2 & 6.8 \\
    n$_{e}$ (10$^{-2}$ cm$^{-3})$ & 2.1$\pm$0.1 & 3.5$\pm$0.2 & $2.24\pm0.05$ & $2.92\pm0.09$ \\
    P (10$^{-10}$ erg\, cm$^{-3})$ & 3.6$\pm$0.3 & 4.1$\pm$0.3 & 2.6$\pm$0.1 & 2.7$\pm$0.2 \\
    $E_{\rm cav}$ = 4pV (10$^{59}$ ergs) & 13.7$\pm$3.3  & 8.1$\pm$1.6 & $19.2\pm3.6$ & $8.5\pm1.9$ \\
    $t_{\rm sonic}$ (10$^{7}$ yr) & 4.6$\pm$0.2&  4.5$\pm$0.1 & $4.3\pm0.4$  & $4.0\pm0.2$\\
    $t_{\rm buoy}$ (10$^{7}$ yr) & 13.1$\pm$0.1 &  9.6$\pm$0.1 & $8.3\pm0.9$ & $7.1\pm0.3$ \\
    $t_{\rm refill}$ (10$^{7}$ yr) & 13.5$\pm$0.3&  11.1$\pm$0.2 & $15.5\pm0.3$ & $12.5\pm0.9$ \\
    $P_{\rm cav}$ (10$^{43}$ erg s$^{-1}$) & 34$\pm$8 & 26.3$\pm$5.0 & $73\pm13$ &  $38\pm8$ \\
   \hline
    \end{tabular}
    \label{energetics}
\end{table}

\begin{figure}
    \centering
    \includegraphics[scale=0.45]{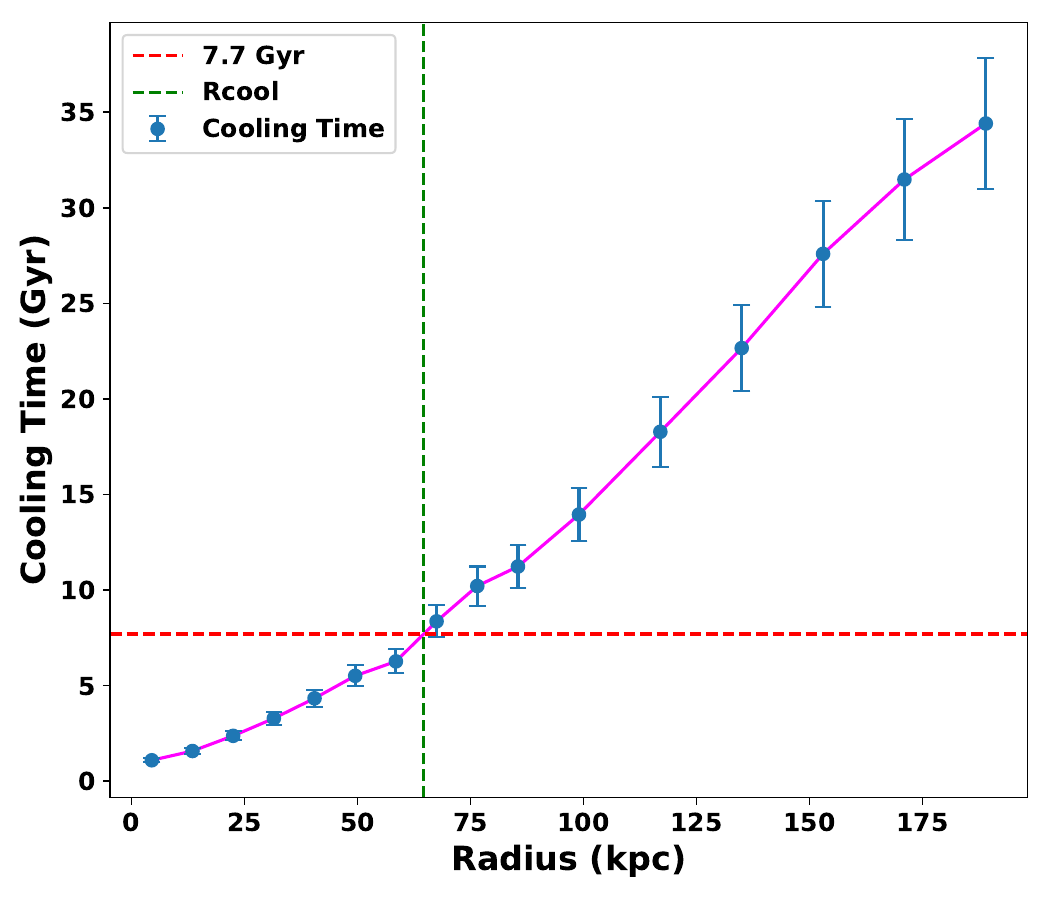}
    \caption{Profile of the cooling time of ICM in the RXCJ1558. The horizontal dashed red line corresponds to the cooling time of 7.7 Gyr.}
    \label{fig:tcool}
\end{figure}

\subsection{Cooling versus heating of the ICM}

In order to study the cooling characteristics of the X-ray emitting ICM in RXCJ1558, we calculated the cooling time profile as shown in Fig.~\ref{fig:tcool} by using electron density and temperature derived from the azimuthally averaged spectral analysis in the 0.5-7.0 keV energy range, based on regions containing at least 3000 counts centered on the X-ray peak. This profile is key to understand the thermal history of the hot gas in the cluster. Each spectrum was fitted individually by \texttt{PROJCT*TBABS*APEC} model. We obtained the cooling radius ($R_{\rm cool}$) (shown by green vertical line) from this profile which is defined by the radius at which the gas cools faster than the Hubble time ($\sim$7.7 Gyr) (shown by red horizontal line). The isobaric cooling time ($t_{\rm cool}$) for the ICM in each annulus was estimated by \cite{1988xrec.book.....S} as:

\begin{equation}
  t_{\rm cool} = 3.5 \times 10^{10} yr
          \biggl(\frac{n_e}{\rm{10^{-3} cm^{-3}}}\biggl)^{-1} \biggl(\frac{T}{\rm{10^{7} K}}\biggl)^{1/2} 
 \label{eq:cooling_time}         
\end{equation}

where, $T$ represents the ICM temperature and $n_{e}$ the electron density. The measured cooling radius is $R_{\mathrm{cool}} = 64.7 \pm 6.0~\mathrm{kpc}$. We also note that the central cooling time of RXCJ1558 lies in the range of 0.9 Gyr $<$ $t_{\rm cool}$ $<$ 7.7 Gyr, indicating a cool-core cluster \citep{2010A&A...513A..37H}. As the cooling luminosity ($L_{\rm cool}$) represents the aggregate radiative power emitted in X-rays, we calculated it within the cooling radius. The resultant cooling luminosity was found to be 1.82 $\pm$ 0.03 $\times$ $10^{44}$ erg s$^{-1}$ \citep[for a detailed calculation see][]{2024JApA...45...23S}.

To examine the `cooling flow' \cite{refId0} have considered two radii (i) cool-core condensation radius ($R_{\rm ccc}$) - the region where the feeding and feedback processes are actively vigorous and (ii) quenched cooling flow radius ($R_{\rm qcf}$) - the region where a quenched cooling inflow could possibly emerge. For RXCJ1558, we adopt the values of $R_{\rm ccc}$ and $R_{\rm qcf}$ as 30 kpc and 62 kpc, respectively \citep{refId0}. We then estimate the corresponding bolometric luminosities for these regions as $L_{\rm ccc}$ = 5.60 $\pm$ 0.07 $\times$ $10^{43}$ erg s$^{-1}$ and $L_{\rm qcf}$ = 1.50 $\pm$ 0.02 $\times$ $10^{44}$ erg s$^{-1}$. For a detailed relationship balance between central AGN heating via cavities and radiative losses of ICM in the RXCJ1558 cluster following \cite{2006ApJ...652..216R}, we plot the power injected by AGN via cavities ($P_{\rm cav}$) versus radiative loss of ICM [$L_{\rm cool}(< r_{\rm cool})$] (Fig.~\ref{added_plot}, left panel). Here, we used \texttt{CADET} estimated value of $P_{\rm cav}$. We also included similar systems available in the literature \citep{2009AIPC.1201..198N,2012MNRAS.421.1360H}. The diagonal lines in the plot, from top to bottom, represent heating rates of 1$pV$, 4$pV$, and 16$pV$ for the equivalence of $E_{\rm cav}=L_{\rm cool}$. In this plot, RXCJ1558 is located closer to the $E_{\rm cav}=1pV$ line than to the $E_{\rm cav}=4pV$ line using $R_{\rm cool}$, $R_{\rm ccc}$, and $R_{\rm qcf}$. This likely reflects possible uncertainties in X-ray cavity power arising from cavity volume estimates, or, alternatively, that the AGN outburst is relatively weak and/or young, such that the cavities have not yet evolved to contain the full enthalpy expected for a relativistic plasma. RXCJ1558 falls within the observed scatter of the relation. Nevertheless, the inferred cavity power is, on average, sufficient to suppress cooling of the ICM, indicating that AGN feedback can be enough to quench radiative cooling in this cluster.

In the right panel of Fig.~\ref{added_plot}, we plotted total mechanical power inferred from the cavity ($P_{\rm cav}$) vs AGN radio power at 1.4 GHz ($P_{\rm 1.4 GHz, AGN}$). This correlation was well studied by \cite{2010ApJ...720.1066C, 2011ApJ...735...11O}. RXCJ1558 follows this relation within the observed scatter, indicating that the radio emission serves as a proxy for the AGN mechanical feedback via X-ray cavities. Here, we have plotted the total mechanical power ($P_{\rm cav}$) calculated from the X-ray analysis by using visual as well as \texttt{CADET} methods vs radio power of the central AGN at 1.4 GHz. The radio power calculated using the flux density measured from sub-band of Band 5 centred at 1.4 GHz. The measured flux density is $288.4\pm28.8$ mJy. The $K$-corrected radio power is $9.4\pm 0.9 \times 10^{40}$ erg s$^{-1}$ calculated at 1.4 GHz using the $\alpha$ $\sim$ - 0.8 (typical for radio galaxies). 
In the plot, RXCJ1558 occupies the position above the best-fit green line, implying that the central radio AGN in RXCJ1558 is capable enough to carve X-ray cavities and quench the cooling flow in RXCJ1558 cluster.

\begin{figure}
\centering
\includegraphics[scale=0.7]{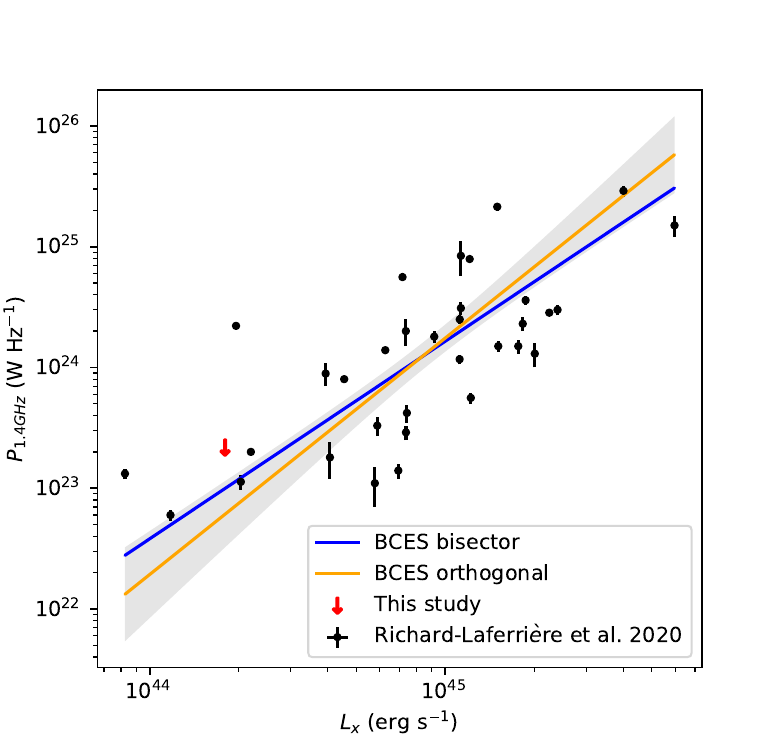}
\caption{Radio mini-halo power ($P_{1.4\,\rm GHz}$) versus X-ray luminosity extracted from 600 kpc region. The plot is reproduced from \citet{Richard_Laferri_re_2020}. RXCJ1558 shows best-fit balance between these two and are marked by red star in the plot.}
\label{RMH-AGN}
\end{figure}

\subsection{Association of sloshing with mini-halo} 
\label{sloshing_and_MH}

In many cool-core galaxy clusters, cold fronts often exhibit sharp edges in X-ray brightness and temperature. These features are typically signatures of sloshing motions in the ICM, triggered by minor off-axis mergers (e.g., \citealt{2001ApJ...562L.153M}; \citealt{2006ApJ...650..102A}). Sloshing not only redistributes the thermal gas, but also induces turbulence and amplifies magnetic fields, creating conditions necessary for re-accelerating thermal particles to relativistic speeds and sustaining diffuse radio emission (\citealt{2008SSRv..134...93F}; \citealt{2015ApJ...801..146Z}).
Our X-ray unsharp, residual and 2D temperature maps of RXCJ1558 reveal a spiral shaped structure, strongly indicative of sloshing activity in the cluster core. This agrees with the findings of the sloshing structure also reported by \cite{2024ApJ...961..134U} for this system. This spiral cold front spatially coincides with the diffuse radio emission detected at Band 3, suggesting that the mini-halo is confined within the sloshing region (see Fig.~\ref{fig:R1558_composite}, left panel). Such a configuration is consistent with MHD simulations by \citet{2015ApJ...801..146Z}, which show that sloshing can generate turbulence and amplify magnetic fields within the core, enabling second-order Fermi re-acceleration of relativistic electrons. The morphology and extent of the observed mini-halo in RXCJ1558 closely resemble the turbulent regions enclosed by cold fronts in these images.

Both the sector-wise projected and deprojected metallicity profiles of RXCJ1558 exhibit a smooth distribution up to a radius of $\sim$72 kpc, i.e., just inside the cold front. At the location of the cold front, we notice a drop in metallicity from 0.63$\pm$0.10 \Zsun\, to 0.35$\pm$0.13 \Zsun\, in the projected profile. Also, the deprojected profile experiences a drop in metallicity, though this remains within the margin of error. We found the spatial coincidence of the cold front, the edge of the mini-halo (see Fig.~\ref{fig:R1558_composite}, left panel), and this sharp metallicity decline suggests a common origin. Comparable metallicity discontinuities have been reported in other clusters and are interpreted as evidence of metal transport in the ICM, driven by sloshing or cold fronts \citep[e.g.,][]{2010MNRAS.405...91S,2010A&A...523A..81D,2011MNRAS.418.2154F,2014A&A...570A.117G}. The disturbed nature of the ICM in RXCJ1558 likely influenced by both AGN feedback and a minor merger event appears to quench strong star formation in the BCG. Indeed, \citet{2022ApJ...940..140C} measured a modest star formation rate of 3.80$\pm$1.14 \Msun yr$^{-1}$. While minor mergers can, under gas-rich conditions, enhance star formation by funneling cold gas into central regions \citep{2014MNRAS.440.2944K}, the situation in BCGs is often different. This system shows evidence of AGN heating, sloshing, and limited cold gas supply in its dense core, as evident from the nuclear quiescent H$\alpha$ emission found in this system using VIMOS observations by \cite{2016MNRAS.460.1758H}. This observation suggests that star formation may remain suppressed. This scenario is consistent with simulations showing that gas sloshing can mix high-entropy gas into the core, raise its entropy, and delay cooling on Gyr timescales \citep{2010ApJ...717..908Z, 2021MNRAS.504.3922C}. Hence, in RXCJ1558, the combination of AGN feedback, stirred-up ICM, and a low cold gas reservoir likely contributes to the observed low but not fully quenched star formation rate in the BCG.

The origin of the relativistic electrons responsible for mini-halo emission remains an open question. A central issue is whether these seed particles are supplied primarily by AGN activity or by ICM. In RXCJ1558, this distinction is particularly important given the presence of a powerful central AGN within a cool core. However, the diffuse radio emission exhibits a markedly asymmetric morphology around the BCG and an unusually steep integrated spectral index ($\alpha_{400{\rm MHz}}^{700{\rm MHz}} < -2.0$, upper limit calculated considering 1$\sigma$ rms of Band 4 map), much steeper than typical AGN-related emission. These characteristics favour a non-AGN origin and instead support a scenario where relativistic electrons are re-accelerated in the ICM.

The mini-halo extends over a projected radius of $\sim$72 kpc, bounded in the south by a cold front located at a nearly identical radius ($\sim$72.4 kpc). This close spatial correspondence may indicate that gas motions which might be associated with sloshing is the main driver of the synchrotron emission. Similar associations between cold fronts and mini-halo edges have been reported in other clusters (e.g. \citealt{2019ApJ...880...70G}; \citealt{2023MNRAS.524.6052R}). In RXCJ1558, further evidence of sloshing comes from the disturbed X-ray morphology, a $\sim$1.2 kpc offset between the BCG and the X-ray peak, central substructures within $\sim$20 kpc, as well as temperature substructures (Fig.~\ref{added_map}). The asymmetric extension of the radio emission, aligned with the cold front, resembles the mini-halo in MS 1455.0+2232 (\citealt{2022MNRAS.512.4210R}), reinforcing the role of sloshing in shaping both morphology and energetics.

From a population perspective, RXCJ1558 is consistent with the empirical correlation between mini-halo radio power at 1.4 GHz and the core-excised X-ray luminosity, when the 1.4 GHz radio power is treated as an upper limit ($P_{\rm 1.4 GHz} \geq 2.47 \times 10^{23}$ W Hz$^{-1}$) for a core excised X-ray luminosity of $L_{\rm X} = 1.80 \times 10^{44}$ erg s$^{-1}$ (\citealt{Richard_Laferri_re_2020}, see Fig.~\ref{RMH-AGN}). This supports its classification as a mini-halo. While the observed asymmetry and steep spectrum could superficially suggest classification ambiguity (similar to Abell 795, \citealt{2024MNRAS.531.4060K}), both traits are consistent with the diversity observed in sloshing-driven mini-halos (e.g. \citealt{2022MNRAS.512.4210R, 2023MNRAS.524.6052R}).

These findings are in agreement with recent theoretical and observational work demonstrating that gas motions associated with the core sloshing can provide sufficient energy to sustain mini-halos (e.g. \citealt{2013ApJ...762...78Z, 2024A&A...686A..82B}). Taken together, the steep spectral index, disturbed X-ray morphology, spatial coincidence with cold fronts, and consistency with established radio - X-ray scaling relations all provide strong evidence that the mini-halo in RXCJ1558 is powered by ICM sloshing, might be driven by turbulence, whose presence could be directly confirmed by upcoming X-ray observatories such as \textit{XRISM} and \textit{NewAthena}. This supports the growing consensus that sloshing is a key mechanism in the origin and evolution of mini-halos in cool-core clusters.


\subsection{Radio mini-halo and AGN feedback connection}
\label{disAGN}
AGN feedback plays a crucial role in heating the ICM. One of the prominent observable evidence of AGN feedback is the X-ray cavities inflated by radio AGN jets \citep{2012ARA&A..50..455F}. Past or present powerful AGN feedback from the central AGN may also be responsible for very extended, steep-spectrum radio emission located outside the cluster core \citep{2004A&A...417....1G,2007A&A...470L..25G}. Such features are interpreted as fossil lobes from previous cycles of AGN activity. A striking example is the Ophiuchus cluster, where \cite{Giacintucci_2020} reported a 500 kpc radio lobe that lies far beyond the central sloshing region that hosts the mini-halo. Similar large-scale fossil lobes have been observed in Abell 2319 \citep{Ichinohe_2021} and in other cool-core cluster such as PSZ1G139.61+24.20, RXJ1720.1+2638 \citep{2018MNRAS.478.2234S, 2019A&A...622A..24S}. We detected two X-ray cavities in the RXCJ1558 cluster orientated in the SE and NW directions, positioned at projected distances of $\sim$36 kpc and $\sim$42 kpc, respectively. Archival VLA 1.4 GHz radio data reported by \cite{2024ApJ...961..134U} as well as our uGMRT Band 5 radio data reveal compact point-like radio emission at the core and an extension of radio emission towards the SE cavity and some further expansion in the east direction evident from our uGMRT data (see Fig.~\ref{new_resid} a). However, it does not fully or partially fill either of the cavities. Previous studies reported many radio filled cavities primarily at 1.4 GHz (e.g., \citealt{2019MNRAS.484.3376L, 2020AJ....160..103P, 2021ApJ...923L..25U, 2025A&A...694A.320H}). In some systems, older generation cavities created by past AGN outbursts are found to be filled with low-frequency radio emission, such as at 610 MHz, 325 MHz \citep{2010ApJ...714..758G, 2019ApJ...870...62P, 2024MNRAS.531.2063L}. \cite{2017MNRAS.466.2054V, 2019MNRAS.485.1981V} for 3C 444 and 3C 320 reported a spatial correlation between X-ray cavities and radio jets/lobes. However, in RXCJ1558, neither the Band 5 nor the Band 3 radio data show evidence of jets or lobes filling the X-ray cavities (see Fig.~\ref{new_resid} a, b). Instead, the cavities are embedded within the broader diffuse emission of the radio mini-halo, with no apparent connection to AGN driven jet or lobe structures. Also, the separation in the location of the cold front and the X-ray cavities suggests that the cold front is unlikely to originate from X-ray cavity expansion (or AGN activity) in the ICM. A similar effort was taken by \cite{2024A&A...686A..82B}, who emphasised the need to distinguish AGN related features (such as cavities and jets) from mini-halo emission. In RXCJ1558, even though a compact AGN like central radio source is present, likely associated with the BCG, it lacks any clear jet or lobe morphology. Instead, it exhibits an amorphous structure characteristic of mini-halos. The morphological and spectral disconnection between the AGN and the radio diffuse emission supports the interpretation that the mini-halo is not a remnant of past or present AGN activity, rather might be arises due to the turbulence driven by ICM gas sloshing within the core of the RXCJ1558 cluster (see Section~\ref{sloshing_and_MH}). However, we cannot entirely rule out the role of AGN feedback in the development of radio diffuse emission such as mini-halo due to shallow uGMRT observations.

\section{Conclusions}
\label{con}
This paper presents an in-depth analysis of 40 ks of \chandra\, data and multi-band uGMRT radio observations on the cool-core galaxy cluster RXCJ1558. We investigated the properties of the X-ray cavities, gas sloshing arm, and cold front confinement of the radio mini-halo in the cluster ICM. The important findings of the present study are summarised as follows.
\begin{itemize}

\item This study confirms that the pair of X-ray cavities exhibit a roughly symmetric and collinear morphology at a projected distance of $\sim$42 kpc in the north-west and $\sim$36 kpc in the south-east of the X-ray peak. 

\item Radio jet activity from the central AGN likely inflated cavities in the ICM as can be visually seen in the residual \chandra\, image. Independent analysis using the \texttt{CADET} tool further supports this suggestion. The enthalpy of the cavities is estimated to be $\sim2.2 \times 10^{60}$ erg and the power is $\sim6.0\pm9.4 \times 10^{44}$ erg s$^{-1}$. We conclude that the power of the observed X-ray cavities is sufficient to compensate for radiative losses, since the cooling luminosity is just $\sim1.5 \times 10^{44}$ erg s$^{-1}$.

\item For our system, cavity volume assessments from visual method and \texttt{CADET} align well and are within a factor of $\sim$2.3 of each other.

\item Observations using the uGMRT Band 3 reveal the presence of diffuse radio emission surrounding the RXCJ1558 cluster. These emissions could potentially be classified as a mini-halo. However, the current shallow observations with uGMRT do not provide enough evidence to confirm the nature of mini-halo. For confirmation, deep radio multiple frequency observations are necessary.

\item This study evidenced a prominent gas sloshing arm and associated cold front at a projected distance of $\sim$72.4 kpc, corresponding to the boundary of the cluster cool-core ($R_{\rm cool} \sim$ $64.7 \pm 6.0~\mathrm{kpc}$), and southern edge of radio mini-halo was found to be spatially coincide with each other, strongly indicating that the radio emitting electrons are confined and re-accelerated by gas sloshing in the ICM. The metallicity gradient is uniform up to the sloshing front (or boundary of the mini-halo), whereas a sharp decrease in metallicity underscores the shared origin of these features.

\item The correlation between radio mini-halo emission, gas sloshing, and AGN activity suggests that the ICM sloshing contributes more to the origin of mini-halo type radio diffuse emission seen in this system.

\end{itemize}

 \section*{Acknowledgments} 
{VSK acknowledges financial support from CSIR, New Delhi (File.No: 09/745(0029)/2021-EMR-I). SKK gratefully acknowledges the financial support of UGC, New Delhi under the Rajiv Gandhi National Fellowship (RGNF) Program. SS acknowledges the Department of Atomic Energy for funding support, under project 12-R\&D-TFR-5.02-0720. NDV and MKP gratefully acknowledge IUCAA, Pune for providing the E-Library facility and their support through Visiting Associate Programme. SIL is supported in part by the National Research Foundation (NRF) of South Africa (CPRR240414214079). Any opinion, finding, and conclusion or recommendation expressed in this material is that of the author(s), and the NRF does not accept any liability in this regard. Data for this work have been obtained from the \chandra\, Data Archive, NASA/IPAC Extragalactic Database (NED), the High Energy Astrophysics Science Archive Research Centre (HEASARC), and Pan-STARRS DR1 imaging data. This work has made use of software packages CIAO and \textit{Sherpa} provided by the \chandra\, X-ray Center. The authors thank the GMRT staff who made observations possible. GMRT is run by the National Centre for Radio Astrophysics (NCRA) of the Tata Institute of Fundamental Research (TIFR).}

 \section*{DATA AVAILABILITY}
 The raw data underlying the RXCJ1558 article are publicly available in the {\it Chandra} (https://cda.harvard.edu/chaser/) and uGMRT archives (https://naps.ncra.tifr.res.in). Analysed data may be made available at a reasonable request to the communicating authors.

\bibliographystyle{mnras}
\bibliography{mybib} 

\appendix
\section{Cavity detection by using \texttt{CADET} tool}
\label{cadet}

\begin{figure*}
    \centering
    \includegraphics[width=0.40\linewidth]{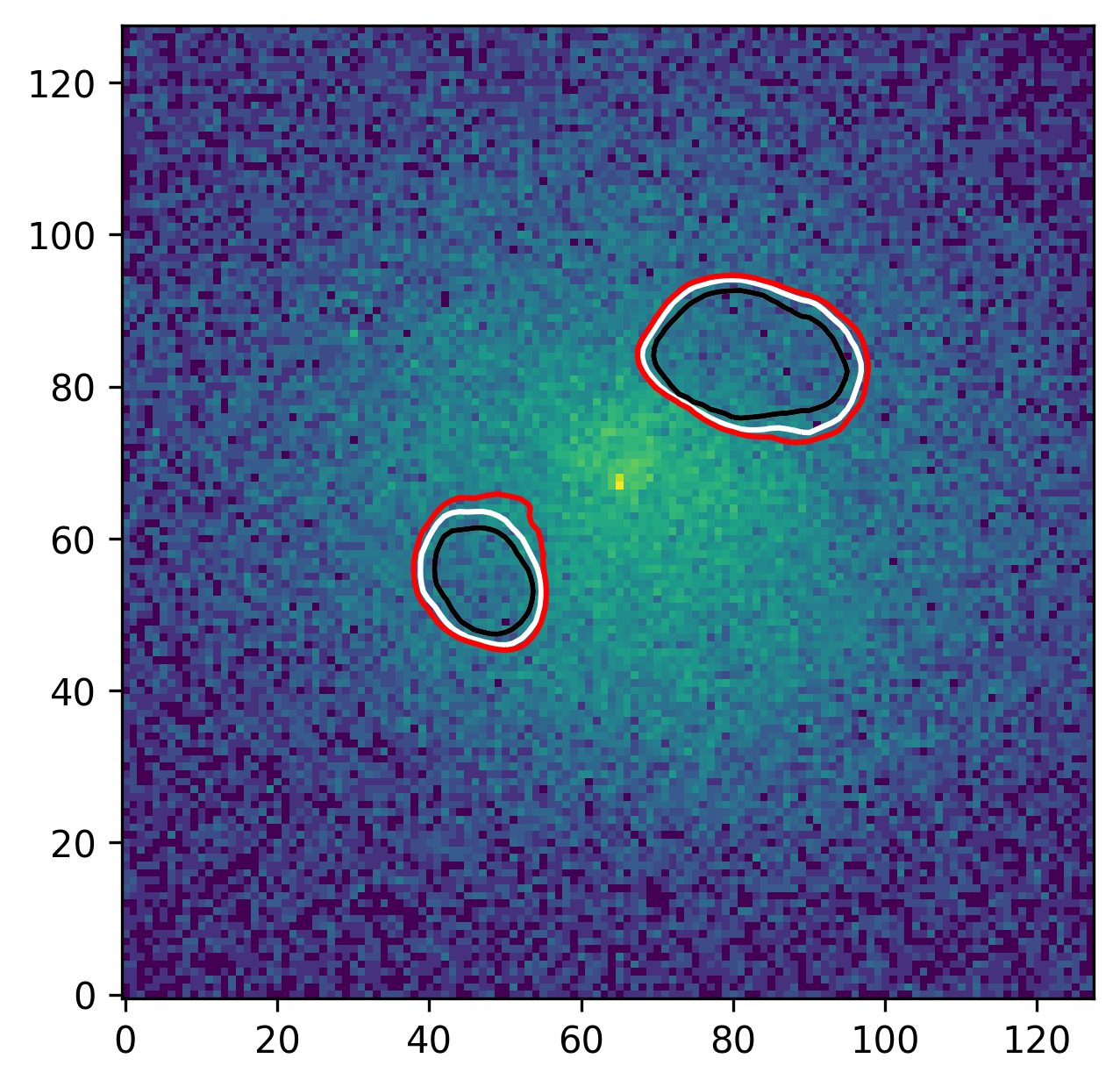}
    \includegraphics[width=0.40\linewidth]{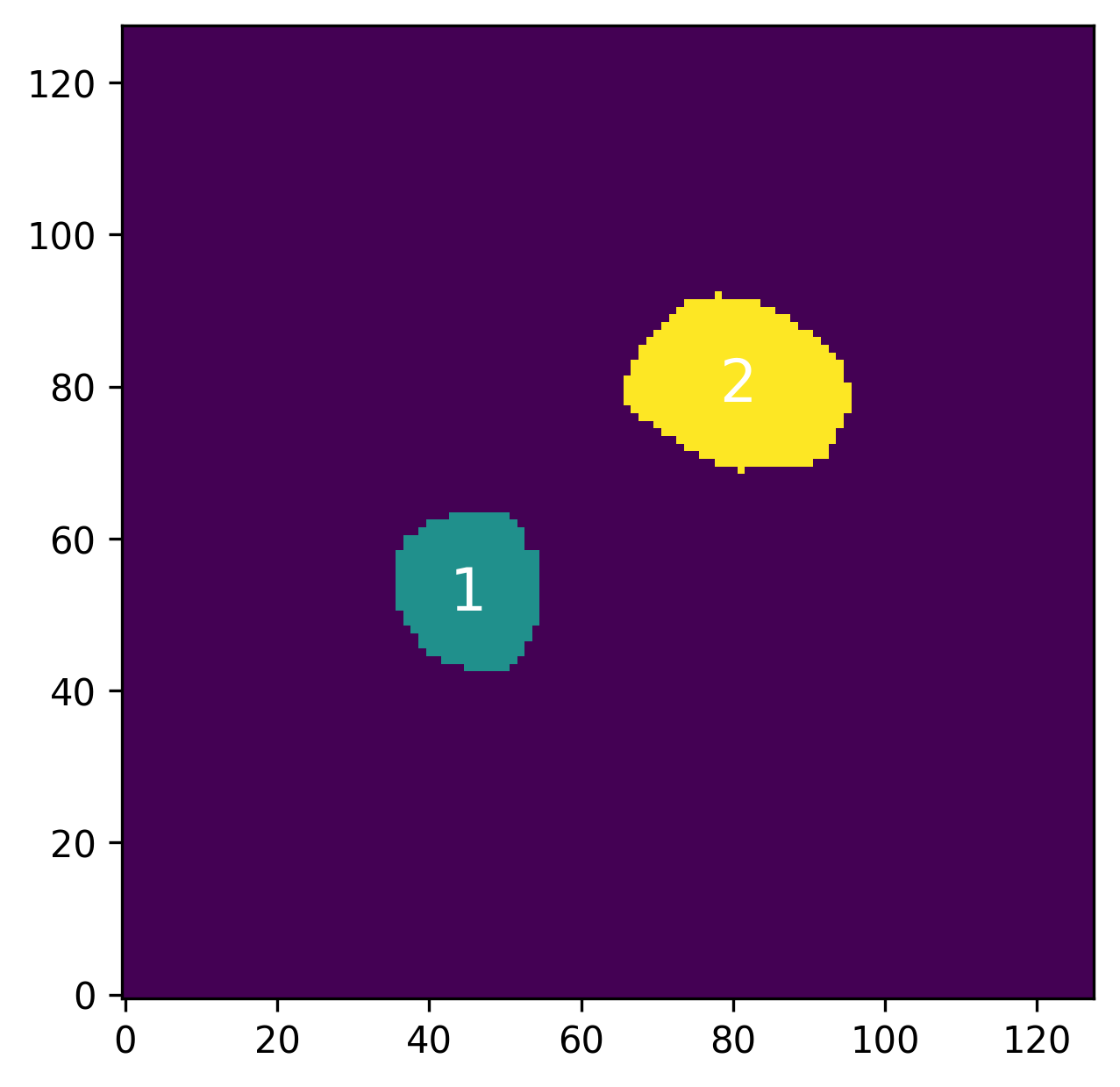}
    \caption{\textit{Left panel} : 0.5-7.0 keV exposure corrected background subtracted 128 $\times$ 128 pixels sized \textit{Chandra} image of RXCJ1558 overlaid with contours of \texttt{CADET} prediction from inner to outward 0.98 (black), 0.65 (white), 0.4 (red) respectively for both X-ray cavities. \textit{Right panel} : Artificially generated images by \texttt{CADET} tool of RXCJ1558 showing two cavities.}
    \label{CADET_fig1}
\end{figure*}

The NW and SE cavities identified in Fig.~\ref{raw} and \ref{new_resid} were confirmed using an alternative method, the CAvity DEtection Tool \citep[\texttt{CADET};][]{2024MNRAS.527.3315P}. This method also helpful to find out morphologies or volume of X-ray cavity by machine learning approach, we applied \texttt{CADET}, which is primarily designed for cavity detection in \chandra\, images. We utilized the broad-band (0.5 − 7.0 keV) image and examined it at scales of 128, 256, 384, 512, and 640 pixels (1 pixel = 0.492 arcsec). The broad-band image was chosen over the 0.5 − 3.0 keV band since \texttt{CADET} based on machine learning algorithm and is specifically tuned for broad-band, with discrimination thresholds adjusted accordingly. Our scale selection was limited by the 128 pixel size of the \texttt{CADET} convolutional neural network. For \chandra\, only 128 and 256 pixel scales were feasible due to the limited field of view. Fig.~\ref{CADET_fig1} shows the resultant image of RXCJ1558 using the \texttt{CADET} tool. We overlaid the contours of \texttt{CADET} predication, from inner to outward 0.98 (black), 0.65 (white), 0.4 (red) respectively for both X-ray cavities. We tried to find out optimal cavity volume threshold. \texttt{CADET} found a considerably strong dependence on the number of counts of the given image. Therefore they binned image by the number of counts into two bins and estimated optimal discrimination thresholds for both bins. For example, images with total counts $<$ 50000 the optimal thresholds are 0.4 and 0.6, and for images with counts $>$ 50000 counts 0.45 and 0.3, respectively \citep[see][for more details]{2024MNRAS.527.3315P}. We utilised the discrimination threshold of 0.4 (for RXCJ1558 counts $\sim$ 85000) to choose for best possible cavity volume, by avoiding over estimation and under estimation of cavity volume. The same volume was used in the Section~\ref{caveng} for cavity energetics calculations by using \texttt{CADET}.


\bsp	

\label{lastpage}
\end{document}